\documentclass[preprint]{aastex}
\usepackage{graphicx}

\newcommand{\yuxi}{YUXI CHEN}
\shorttitle{CMD and Morphology of SDSS ULIRGs}
\shortauthors{Chen et al.}

\begin{document}

\title{Color Magnitude Relation and Morphology of Low-Redshift ULIRGs in SDSS}

\normalsize \author{\yuxi \altaffilmark{1}, 
JAMES D. LOWENTHAL\altaffilmark{2}, and MIN S. YUN\altaffilmark{1}}
\affil{$^1$ Department of Astronomy, University of Massachusetts, Amherst, MA 01003\\  
      $^2$ Department of Astronomy, Smith College, Northampton, MA 01063}
\altaffiltext{1}{Email: yxchen@astro.umass.edu}

\begin{abstract}

We present color-magnitude and morphological analysis of 54 low-redshift ultraluminous
infrared galaxies (ULIRGs; $0.018<z<0.265$ with $z_{median}=0.151$),
a subset of the IRAS 1Jy sample (Kim \& Sanders, 1998), in the Sloan Digital Sky Survey (SDSS).
The ULIRGs are both bright and blue: they are on average
1 magnitude brighter in $M_{^{0.1}r}$ than the SDSS galaxies within the same redshift range,
and 0.2 magnitude bluer in $^{0.1}g-^{0.1}r$. They form a group in the color-magnitude diagram distinct
from both the red sequence and the blue cloud formed by the SDSS galaxies: 24 out of the 52 unsaturated
objects ($\sim46\%$) lie outside the 90\% level number density contour of the SDSS galaxies.
The majority (47, or $\sim 87\%$) have the colors typical of the blue cloud, and only 4 ($\sim 6\%$) sources are
located in the red sequence.
While ULIRGs are popularly thought to be precursors to a QSO phase, we find few (3 or $\sim6\%$) 
in the ``green valley'' where the majority of the X-ray and IR selected AGNs are found.  
Moreover, none of the AGN-host ULIRGs are found in the green valley.
For the 14 previously spectroscopic identified AGNs ($\sim28\%$), we perform PSF subtractions and find that on average the central point
sources contribute less than one third to the total luminosity, and that their high optical luminosities
and overall blue colors are apparently the result of star formation activity of the host galaxies. 
Visual inspection of the SDSS images reveals a wide range of  
morphologies including many close pairs, tidal tails,  
and otherwise disturbed profiles, in strong support of previous
studies and the general view of ULIRGs as major mergers of gas-rich  
disk galaxies. A detailed morphology analysis using
$Gini$ and $M_{20}$ coefficients shows that slightly less than one half ($\sim42\%$ in $g$ band) of the ULIRGs are
located in the merger region defined by morphology studies of local galaxies,
while the remaining sources are located in the region of late-type and irregular galaxies.
The heterogeneous distribution of ULIRGs in the $G-M_{20}$ space is qualitatively consistent with the
results found by numerical simulations of disk-disk mergers, and our study also shows
that the measured morphological parameters are systematically affected by the signal-to-noise
ratio and thus the merging galaxies can appear in various regions of the
$G-M_{20}$ parameter space.
We briefly discuss the origins of the uncertainties and note that the morphology measurements
should be implemented with caution for low physical resolution images.
In general, our results reinforce the view that ULIRGs contain young stellar populations and are mergers in progress, 
but we do not observe the concentration of ULIRGs/AGN in the green valley as found by other studies. 
Our study provides a uniform comparison sample for studying dusty starbursts at
higher redshifts such as Spitzer MIPS 24$\mu$m-selected ULIRGs at $z=1\sim2$ or submillimeter galaxies.

\end{abstract}


\section{INTRODUCTION}

Ultraluminous infrared galaxies (ULIRGs; $L_{IR}>10^{12}\rm L_{\odot}$) were first 
discovered in a large number by {\it Infrared Astronomical Satellite} (IRAS) 
two decades ago (Rowan-Robinson et al. 1984; Soifer et al. 1984). These objects are among the most 
luminous sources in the universe, with most of their energy radiated in the infrared. 
Although the powering sources of ULIRGs remain uncertain, there is evidence that 
both AGNs and dusty starbursts contribute to their high bolometric 
luminosities (see Sanders \& Mirabel 1996; Lonsdale, Farrah, \& Smith 2006 for reviews). The majority of local ULIRGs 
are mergers with tidal features commonly seen (e.g. Sanders et al. 1988; Melnick \& Mirabel 1990; Murphy et al. 1996), 
and distant ULIRGs are the possible progenitors of today's massive ellipticals (e.g. Barnes \& Hernquist 1996).

Recent large optical photometric surveys show that the color distribution of galaxies are bimodal
(Strateva et al. 2001; Hogg et al. 2003; 
Blanton et al. 2003c; Bell et al. 2004a,b). Color-magnitude diagram reveals a red sequence of early type galaxies 
with little star formation activity, and a blue cloud of late type 
galaxies that form stars. The characteristic number density in the standard luminosity function of 
Schechter (1976), ${\phi_{*}}$, has doubled for the red sequence galaxies since $z\sim1$, 
but hardly changes for the blue cloud, implying that the number and stellar mass of the red sequence 
galaxies have been gradually built up, whereas those of the blue cloud galaxies remain nearly constant 
(e.g., Bell et al. 2004b; Faber et al. 2007). 
One scenario of building up the most massive galaxies in the 
red sequence is major merging of disk galaxies (Bell et al. 2004b), and thus galaxies in the 
``green valley'', the region between the red sequence and blue cloud in the 
color-magnitude diagram, may represent this transitional type. As ULIRGs are proposed 
to be disk-disk mergers that will finally evolve to massive ellipticals,
studying the color-magnitude relation of ULIRGs may shed light on the evolutionary scenarios of 
building up massive ellipticals.

Galaxy morphologies indicate the distribution of baryonic matter in the galaxies, and are therefore
tracers of its formation and evolution.
The majority of local ULIRGs are merging systems in which disturbed morphologies such as tidal features are commonly observed.
Recent deep sky surveys have generated enormous amounts of galaxy imaging data, and automated morphology
measurement methods are required to analyze those large datasets with much higher efficiencies 
than visual inspections provide. There are two quantitative approaches to describing galaxy morphology: 
parametric methods and non-parametric methods. Parametric methods fit the light distribution 
of a galaxy to pre-defined formulae, such as Sersic index fitting (Blanton et al. 2003c) and 
bulge-to-disk ratio (Peng et al. 2002; Simard et al. 2002), and are usually inadequate 
to describe the morphologies of complex systems. Non-parametric methods, 
such as the concentration, asymmetry, and clumpiness (CAS) system 
(e.g. Isserstedt \& Schindler 1986; Abraham et al. 1994; Wu et al. 1999; 
Conselice et al. 2000; Conselice 2003), do not rely on pre-defined functions,
and may be the better methods for describing complex systems. Abraham et al. (2003) introduced the 
$Gini$ coefficient ($G$), which describes the relative intensity distributions of a galaxy. The $Gini$ coefficient
is correlated with concentration and increases with the fraction of light in compact components. 
Lotz, Primack, \& Madau (2004; hereafter LPM04) introduced $M_{20}$, which is the second moment of the brightest 20\% of galaxy
flux. $M_{20}$ is sensitive to spatial distributions of off-axis clumps. $G$ and $M_{20}$ thus should be 
sensitive to the presence of merger features and indeed LPM04 found that local ULIRGs are well separated from 
normal galaxies in $G-M_{20}$ space, in the sense that ULIRGs have higher $G$ and $M_{20}$ values than normal galaxies.

In this paper we present the results of color-magnitude analysis and $G$ and $M_{20}$ morphological analysis
of 54 ULIRGs from the $IRAS$ 1Jy ULIRG sample (Kim 1995; Kim \& Sanders, 1998) that were observed by Data Release 5
of Sloan Digital Sky Survey. The goals of this paper are to place ULIRGs 
in the context of normal galaxies by comparing their color-magnitude relations 
and morphologies with those of optically-selected SDSS galaxies, and to construct a low-redshift comparison sample 
for studying high-redshift dusty starbursts, 
such as Spitzer-identified ULIRGs and submillimeter galaxies (SMGs). 
The $IRAS$ 1Jy ULIRG sample is a complete flux-limited sample from the IRAS Faint Source Catalog 
(Moshir et el. 1990)
with $S_{60\mu m}>1 \rm Jy$ and $\delta>-40^{\circ}, |b|>30^{\circ}$ (Kim \& Sanders 1998).
The sample contains 118 sources with a redshift range between 0.018
and 0.268 and a median redshift of 0.145. The infrared luminosities range between $10^{12.00}$ and $10^{12.90}\rm L_{\odot}$, 
with a median of $10^{12.19}\rm L_{\odot}$.  
They are the most IR luminous galaxies in the low redshift universe, and their uniform selection and completeness 
make them ideal for a statistical analysis.
Almost all ULIRGs in the sample show visual signs of interactions or mergers, but most are at advanced stages, 
and the median R-band luminosity is $\sim2L_*$ (Kim, Veilleux, \& Sanders 2002, hereafter KVS02; 
Veilleux, Kim, \& Sanders 2002, hereafter VKS02).
Spectroscopic study showed that the Seyfert fraction increases with infrared luminosity, and $\sim50\%$ of the galaxies with 
$L_{IR}>10^{12.3}L_{\odot}$ present Seyfert characteristics. These AGN-powered ULIRGs have
bolometric luminosities and near-IR spectra reminiscent of the local quasars. Another $30\%$ of the whole sample are classified as 
H{\sc ii}-region and $\sim40\%$ are classified as LINERs, in both of which the energy sources are thought to be 
stellar origin from recent starbursts ($\leq$ few times $10^7$ yr) or shocks 
(Veilleux, Kim \& Sanders 1999, hereafter VKS99; Kim, Veilleux, and Sanders 1998)
Near-IR study using 2MASS showed that Seyfert galaxies among the 1Jy sample have much redder colors 
and steeper spectral indices in the near-infrared than the rest, 
which indicates powerful AGNs as the energy sources (Chen \& Zhang 2006). 
Recent $^{12}$CO ($J=1\rightarrow 0$) observations of 17 ULIRGs in the 1Jy Sample, along with 12 other local ULIRGs 
showed that their large IR luminosity and gas mass to FIR luminosity ratios are consistent with a model where most of their 
luminosity is powered by a brief surge in star formation rate associated with the rapid gas inflow during the merger phase, 
although there is also evidence for powerful AGNs with extreme $L_{FIR}/L_{CO}^{\prime}$ among a subset and they are possibly 
transitioning to a quasar phase (Chung et al. 2009).
In general, previous studies suggest an evolutionary sequence 
of ULIRGs in which ULIRGs start with cool infrared colors ($f_{25\mu m}/f_{60\mu m}<0.2$), 
go through a warm ULIRG state ($f_{25\mu m}/f_{60\mu m}>0.2$), and finally evolve to quasars (Sanders 1988; VKS02).   
 
In \S\ref{sample_selection} we present the sample selection. The image analysis is presented 
in \S\ref{image_analysis}. We present the color magnitude relation of the ULIRGs in \S\ref{cmr}, 
their morphological analysis in \S\ref{morphology}, and the summary in \S\ref{summary}.
We use a $\Lambda$CDM cosmology with $H_0=70\rm kms^{-1}Mpc^{-1}$, $\Omega_m=0.3$ and $\Omega_{\Lambda}=0.7$ 
throughout the paper, except for the absolute magnitudes, for which we use $h=1$, 
where $h=H_0/100\rm kms^{-1}Mpc^{-1}$, for a direct comparison with existing studies.

\section{SAMPLE SELECTION} 
\label{sample_selection}

The sample consists of all the sources from the $IRAS$ 1Jy ULIRG sample 
(Kim 1995; Kim \& Sanders 1998) that have been imaged in the fifth release 
of the Sloan Digital Sky Survey (SDSS DR5; Adelman-McCarthy et al. 2007). 

The SDSS is the largest homogeneous, wide-area multiwavelength sky survey available 
for studying galaxy morphology and color-magnitude distributions in the low redshift universe.
We downloaded SDSS DR5 images at the coordinate position of each source in the 1Jy ULIRG sample, 
and did a one-by-one comparison with published optical and near-IR images 
(KVS02). This yielded a final sample of 54 ULIRGs identified in SDSS. 
The 54 sources span a redshift range between 0.018 to 0.265, with a median of 0.151, and an infrared 
luminosity range between  $10^{12.00}$ and $10^{12.76}\rm L_{\odot}$ with a median of $10^{12.23}\rm L_{\odot}$. 
The mean values of our sample are very close to those in the whole 1Jy sample. We perform
Kolmogorov-Smirnov (K-S) tests between our sample and the whole 1Jy sample for the distributions of redshift, 
infrared luminosity, and infrared colors ($F_{25\mu m}/F_{60\mu m}$), and the probabilities that the two
samples are drawn from the same distribution are 80\%, 97\%, and 99\%, for redshift, infrared luminosity, 
and infrared colors, respectively. Thus we conclude that our sample is a good representative subset of the whole
1Jy sample. 

The redshifts and the FIR luminosities of the sample are presented in the second and third column in 
Table \ref{tab:sample}, respectively. There are 14 ULIRGs harboring one or more AGNs 
according to optical and near-IR spectroscopic study (VKS99 and references therein), 
and they are identified accordingly in Table \ref{tab:sample}. The AGN sample includes all galaxies with 
Seyfert activity but excludes all LINERs, which are thought to be low-energy AGNs (e.g. Krolik 1998).
Earlier studies of the 1Jy ULIRG sample concluded that the AGN fraction increases with IR luminosity (VKS99; VKS02).

The comparison sample we choose for studying the color-magnitude relation is the 436,762 galaxies 
in the DR4 of the NYU Value Added Galaxy Catalog (NYU-VAGC, DR4; Blanton et al. 2005) with spectroscopic redshifts in the same redshift range
of our ULIRGs. For studying the morphological parameters, the comparison sample we selected consists of the galaxies that 
appear on the same SDSS images of the ULIRGs. The size of each image is $13.51\arcmin \times 8.98 \arcmin$, 
and there are 1215, 2059, 2019, and 484 comparison galaxies, in the g, r, i, and z band, respectively. 
We will discuss the selection method in more detail in \S\ref{morph.calc}.

\section{IMAGE ANALYSIS}
\label{image_analysis}

We obtained pipeline-reduced clean images ({\it fpC*.fit}) from the SDSS DR5 database. 
We adopted the skylevels in the observation auxiliary files ({\it tsField*.fit}) and subtracted them from the images.
Each image was checked to ensure the soft skyvalue in the header is correct; for incorrect sky values, we 
estimated the sky levels based on the statistics of multiple small empty-sky regions on the same image. 

\begin{figure}
\includegraphics[width=6in]{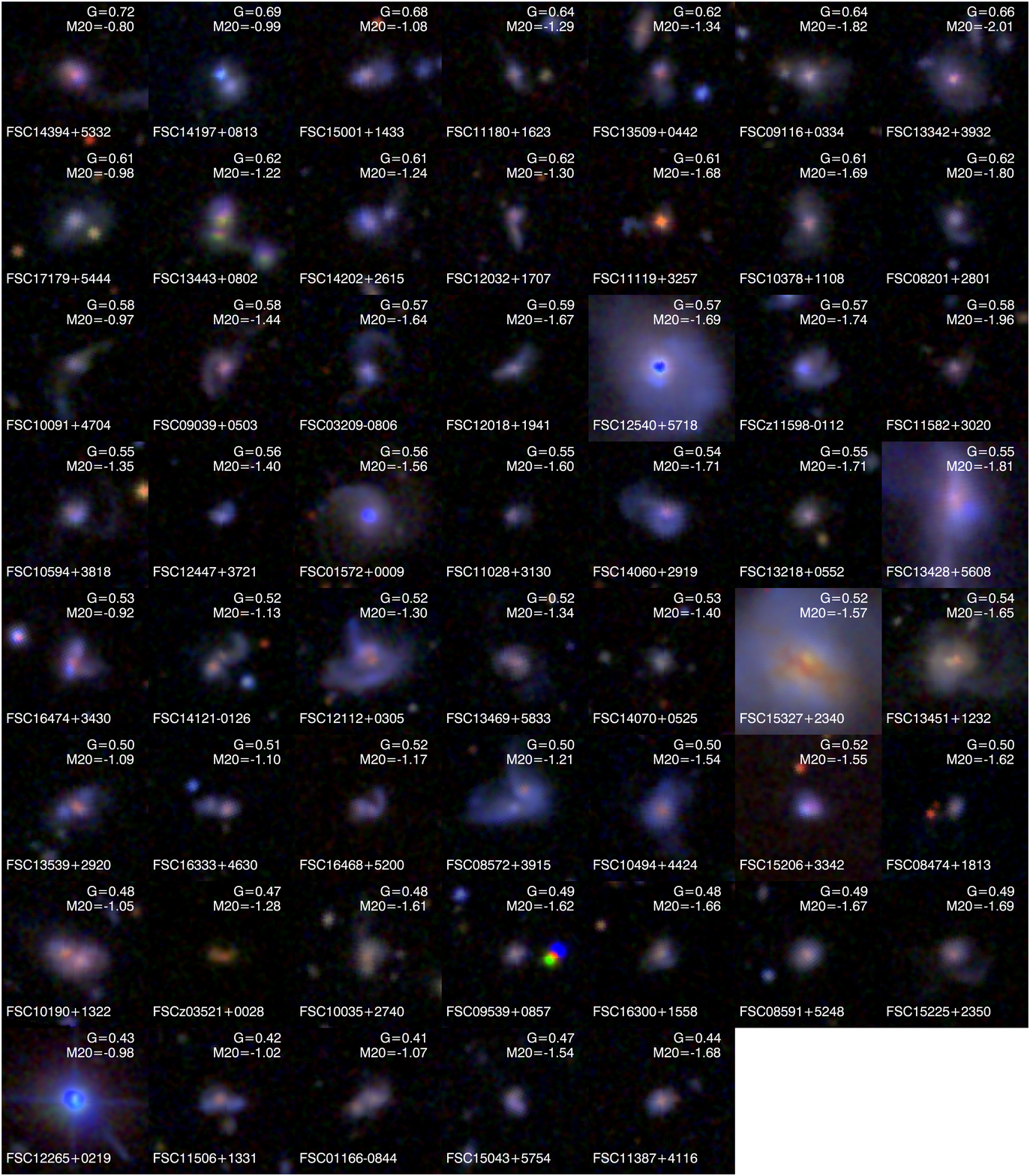}
\caption{Mosaic of RGB color images of all 54 sources. North is up and east is left for each image. 
Each image is 100 pixels across, or $\sim40\arcsec$ at SDSS's
pixel scale, and centered on the source. The images are arranged according to their
$g-$band $G$ and $M_{20}$ (see \S\ref{morphology}), so that for each column $G$ increases
towards the top, and for each row $M_{20}$ increases towards the left. The $G$ and $M_{20}$
are printed on each image. A clear trend is seen: more disturbed and multiple nuclei sources
are located towards the upper left corner of the mosaic.}
\label{fig1}
\end{figure}

A mosaic of RGB color images of all 54 sources
is shown in Fig. \ref{fig1}. (Refer to the electronic version for the color image). 
Each image is 100 pixels across, or $\sim40\arcsec$ at SDSS's pixel scale,
and centered on the source. 
The color images were prepared using the method described by Lupton et al. (2004); 
$g-$, $r-$, and $i-$band background subtracted images as blue, green, and 
red images, respectively. The images were smoothed with a median window of 2
pixels wide, aligned using $r-$band images, and rotated so that north is up.
The relative scales of red, green, and blue images were set to 1:1:2, in order 
to emphasize the short wavelength radiation from young stellar populations. 
The first visual impression  
is that the majority of the galaxies 
show disturbed/irregular morphologies, as well as strong color gradients and discontinuities. 
Note that there is a smeared three color source on the image of FSC09539+0857. We checked the SDSS Moving 
Object Catalog (MOC, the third release; Ivezic et al. 2002) and found that it is an asteroid with Unique SDSS moving object ID s1d059.

The images are arranged according to the $g-$band $G$ and $M_{20}$ coefficients of each source (see {\S\ref{morphology}}), 
so that for each column $G$ increases towards the top, and for each row $M_{20}$ increases towards the left. 
The coefficients were calculated using the method described by LPM04 (see \S\ref{morph.calc}) and are printed on each image. 
Although the images are arranged only relatively in the $G-M_{20}$ parameter space, a clear trend
is seen: more disturbed and multiple nuclei sources have larger $G$ and $M_{20}$ and are 
located toward the upper left corner of the mosaic, and the less disturbed sources, with smoother and more symmetric morphology, 
have smaller $G$ and $M_{20}$ and tend to be located toward the lower right corner.  
Visual inspection of these images also shows a wide range of
morphologies including many close pairs, tidal tails, and otherwise disturbed profiles, in strong support of previous
studies and the general view of ULIRGs as major mergers of gas-rich disk galaxies. 

We computed the magnitude within an elliptical aperture whose semi-major axis
is 2 Petrosian radii (Petrosian 1976) for each source.
The Petrosian radius is defined as the radius of a circular aperture on which
the mean surface brightness is equal to 20\% of the average surface brightness
contained inside the aperture. Using an aperture as large as 2 Petrosian radii
ensures that most low-surface-brightness features are included in the photometry,
and the SDSS Petrosian magnitudes we use to comparison are also obtained within 
a 2 Petrosian radius aperture for each source (Blanton et al. 2001; Yasuda et al. 2001).
For each source, we align the five images using 
the $r$-band image as the reference. Bilinear interpolation is used during the transformation of each image, and flux
is estimated to be conserved within the 0.05\% level. We run $SExtractor$ (Bertin \& Arnouts 1996) on the $r$-band image 
and obtain the aperture whose semi-major axis is 2 Petrosian radii, and perform the photometry for all the five bands using the same aperture. 
Some sources have disconnected multiple components detected by $SExtractor$, 
with each component confirmed to be a part of the ULIRG by its redshift based on previous studies (KVS02). 
For these sources, if the components do not have overlapping apertures, we adopt the total flux from multiple 
components for photometry; if they have overlapping apertures, we fit the overlapping region with an ellipse and measure 
the flux within the ellipse, then subtract it from the total flux to obtain the corrected flux of the entire source.
The airmass and the photometric zero points are obtained from the auxiliary files ({\it tsField*.fit}) and applied when calculating the 
magnitudes. The SDSS magnitudes were then converted to AB magnitudes using the AB-SDSS magnitudes corrections. 

We find that for the ULIRGs the independently measured apertures are not identical in all five bands. 
This is because many ULIRGs are extended and amorphous, and the extensions of low surface brightness
features can be different at different wavelengths. The mean relative difference of the Petrosian radii measured 
in $g$ and $r$ band, $\langle{(r_{P,g}-r_{P,r})/r_{P,r}}\rangle $, is $\sim8\%$.
We perform the $g-$band photometry using the apertures measured in the $g$ band, 
and find that the measured magnitudes are on average 0.03 mag more luminous than the magnitudes measured 
using the $r-$band apertures, and the standard deviation of the difference is 0.10 mag. 
Because we use $r-$band apertures to perform the photometry, we adopt this standard deviation as 
a typical error of the $^{0.1}g-^{0.1}r$ color for the ULIRGs.
 
Our main purpose is to locate the positions of the ULIRGs 
in the color-magnitude diagram, and compare them with the existing survey data. We 
therefore adopt the k-corrections to SDSS bandpasses shifted to $z=0.1$, 
the median SDSS galaxy redshift, employing the {\it kcorrect v4.2} package developed by Blanton et al. (2003a).
Each shifted band is denoted by a superscript on the left, e.g., $^{0.1}g$ denotes the SDSS $g$ band shifted to $z=0.1$,
and $^{0.1}g-^{0.1}r$ denotes the k-corrected color between two shifted bands, $^{0.1}g$ and $^{0.1}r$.
Galactic extinction is corrected using the SFD98 dust map (Schlegel, Finkbeiner \& Davis 1998). 
We derive the absolute magnitude by applying the distance modulus to the k-corrected apparent magnitude. 
We present the photometry results in Table 1.

\section{COLOR-MAGNITUDE RELATION}
\label{cmr}
\subsection{Color and Magnitude of SDSS Comparison Sample}

In Fig. 2 we plot the $^{0.1}g-^{0.1}r$ color against k-corrected $^{0.1}r$-band absolute magnitude $M_{^{0.1}r}$ for all
comparison sample galaxies, which consists of the 436,762 galaxies from NYU SDSS Value Added Catalog (DR4) 
within the same redshift range. Individual galaxies are shown as small dots while the contours represent 
the galaxy number density. The bimodal distribution of the SDSS comparison galaxies is clearly shown 
in the color-magnitude diagram where the red sequence and the blue cloud form two separate groups. 
We adopt the empirical definition of the red sequence from Weinmann et al. (2006), 
such that the red-sequence galaxies follow the color-magnitude relations,
\begin{equation}
^{0.1}g-^{0.1}r > 0.7 - 0.033(M_{^{0.1}r}-5logh+16.5)
\end{equation}
This relation is shown in Fig. 2 where the red sequence lies above the upper straight line. The centroids of the red sequence and 
the blue cloud are separated by $\sim 0.3$ in $^{0.1}g-^{0.1}r$, and between the two populations there is a relatively 
low density region called the ``green valley''. 
We define the green valley as a strip below the red sequence with a width of 0.1 in  $^{0.1}g-^{0.1}r$, 
shown as the region between the two straight lines in Fig. 2. 
The fiducial width of 0.1 is chosen to be one third of the color 
difference between the centroids of the red sequence and the blue cloud, and is also roughly equal to the color difference between the centroid 
and the blue limit of the red sequence (the upper straight line). The low density region between the red sequence and the blue cloud 
is wider at fainter magnitudes and narrower at brighter magnitudes, and our simple cut may miss some green-valley galaxies 
at fainter magnitudes but include more blue-cloud galaxies at brighter magnitudes. 
However, this will not affect our main results as we discuss later.

\subsection{ULIRGs Are Optically Bright and Blue}

Table \ref{statistics} lists the statistics of the colors and magnitudes of the ULIRGs and SDSS comparison sample and their subsamples. 
We summarize below those optical characteristics of the ULIRGs and their subsamples (AGN and non-AGN) and compare them with 
those of the SDSS comparison sample.

The first obvious trend is that the ULIRGs are very luminous in the optical.
The median k-corrected absolute magnitude is $M_{^{0.1}r} = -21.4$, or equivalently 2.5 $L_*$ given $M_{^{0.1}r}=-20.44$ 
for an $L_*$ galaxy (Blanton 2003b). This is consistent with the median of $R-$band absolute magnitudes for the whole 1 Jy sample 
($\sim 2L_*$; KVS02). For comparison the median k-corrected absolute magnitude of the SDSS galaxies 
is $M_{^{0.1}r}=-20.4$, which is 1 magnitude fainter than that of the ULIRGs. 
The absolute magnitudes of the ULIRGs range from $M_{^{0.1}r} = -19.78$ to $M_{^{0.1}r} = -23.23$ 
(excluding the two saturated sources), a luminosity range of a factor of 25. Fifty out of the 54 ULIRGs 
($\sim93\%$; including the two saturated sources) are more luminous than the median k-corrected $r-$band absolute magnitude of the 
SDSS comparison sample. The difference in magnitude distribution between the ULIRGs and the 
SDSS comparison sample is further illustrated in Fig. \ref{maghist}. 
The ULIRG sample has a much narrower distribution than that of the 
comparison sample. The ULIRGs are on average 1 magnitude brighter than that of the entire comparison sample (see panel(a)) as well as 
the individual subsample of the red sequence, the green valley, and the blue cloud (panel (c)).
At the highest luminosity bins the ULIRG sample outnumbers all three SDSS sub-samples in fraction 
($M_{^{0.1}r}-5logh<-21$; panel (c)).

Another striking trend is that the ULIRGs are very blue, with only a few exceptions.
Their $^{0.1}g-^{0.1}r$ colors span a wide range, from $0.04$ to $1.5$, 
and the majority (87\%) of the ULIRGs have optical colors as blue as those of the blue-cloud galaxies.  
The median $^{0.1}g-^{0.1}r$ color of the ULIRGs is 0.58 and very close to that of the blue cloud (0.55). 
The median $^{0.1}g-^{0.1}r$ color of the red sequence and green valley are 0.95 and 0.78, respectively. 
The color distributions of the ULIRGs and the SDSS galaxies are further 
illustrated in Fig. \ref{colorhist}(a) and (c). 
Considering galaxy color as a function of galalxy luminosity, we limit the $r$-band absolute magnitudes
of the SDSS comparison sample to the same magnitude range as that of the ULIRGs ($-19.8<M_{^{0.1}r}-5logh<-23.2$). 
The peak of the ULIRG distribution is $\sim0.3$ mag bluer than that of the SDSS sample within the same $r$-band magnitude range, 
but very similar to that of the blue cloud within the same $r$-band magnitude range ($^{0.1}g-^{0.1}r\sim0.6$). 
The ULIRGs have a much larger fraction in the bluer bins ($0.5<^{0.1}g-^{0.1}r<0.7$) than the comparison sample,
but a much lower fraction in the redder bins ($0.7<^{0.1}g-^{0.1}r<1.0$), except for the reddest bins 
($^{0.1}g-^{0.1}r>1.2$) and the bluest bins ($^{0.1}g-^{0.1}r<0.1$). The ULIRG sample outnumbers the blue-cloud 
galaxies in fraction at blue colors ($^{0.1}g-^{0.1}r<0.7$), has a similar fraction compared to the green-valley galaxies
at green colors ($0.7<^{0.1}g-^{0.1}r<0.9$), and has a much lower fraction than the red-sequence galaxies 
at red colors ($^{0.1}g-^{0.1}r>0.9$).


The blue colors of the ULIRGs seem inconsistent with the working hypothesis that ULIRGs are dusty, massive systems.
However, the patchy mix of blue and red regions in the optical color images (see 
color version of Fig. 1) suggests that the dust extinction is patchy, and the extended distribution of young, 
blue starts dominate the overall color of these galaxies.
Veilleux, Sanders \& Kim  (1999) have discussed the blue optical continuum colors of the ULIRGs and 
suggested that the stellar light suffer less dust extinction than the emission-line gas. 
Our ULIRGs would appear even more luminous and bluer 
in the optical if corrected for the internal extinction.

The AGN ULIRGs are among the most luminous sources in the ULIRG sample, with a median $M_{^{0.1}r} = -21.9$, 
0.8 magnitude more luminous than the median of the non-AGN ULIRGs. 
As shown in Fig. \ref{maghist}(b) the AGN ULIRGs are more concentrated
toward high luminosities and outnumber the non-AGN ULIRGs at the highest luminosity bin ($M_{^{0.1}r}-5logh<-23$).
They also have a much flatter color distribution than the non-AGN ULIRGs (see Fig. \ref{colorhist}(b)). 
There is little overlap between the AGN ULIRGs and the green-valley galaxies in both magnitude 
distribution (see Fig. \ref{maghist}(d)) and color distribution (see Fig. \ref{colorhist}(d)). 

\subsection{ULIRGs Are Scarce in the Green Valley}

We overlay the colors and magnitudes of the ULIRG sample on top of the SDSS contours in Fig. 2, with different symbols
representing ULIRGs known to host an AGN (AGN ULIRGs, circles) and H{\sc ii}-region like and LINER galaxies (non-AGN ULIRGs, stars).
The images of FSC12540+5708 and FSC12265+0219 are saturated in the $g$ and $r$ band, 
and their magnitudes are shown as upper limits. The error bar shows $\sigma(g_{P,g}-g_{P,r})$, 
the standard deviation of difference in $g-$band magnitude measured between $g-$ and $r-$band apertures 
(\S\ref{image_analysis}).

The ULIRGs form a distinct group in the color-magnitude diagram: 24 out of the 52 unsaturated ULIRGs ($\sim46\%$)
lie outside the 90\% level contour of the SDSS galaxies in the same redshift range. 
Strikingly, only 3 ($6^{+6}_{-4}\%$\footnotemark) lie in the green valley, and none hosts an AGN. 
The majority of the ULIRGs (47 out of 54, or $87^{+7}_{-11}\%$) are located below the bottom straight line 
that defines the red limit of the blue cloud, and 4 ($7^{+5}_{-6}\%$) are located above the upper 
straight line that defines the blue limit of the red sequence.
\footnotetext[1]{The uncertainty on the fraction of sources in any given region of two-dimensional parameter spaces, 
e.g. the fraction of red-sequence galaxies in the color-magnitude relations, or the fraction of mergers in the $G-M_{20}$ plot, 
is calculated from the possible miscounted number of sources in this region given typical errors in both parameters.}

In the local universe most ULIRGs are major mergers of late-type galaxies, and are proposed to be the precursors of massive 
ellipticals (e.g. Toomre \& Toomre 1972; Toomre 1977; Mihos \& Hernquist 1994, 1996; Barnes \& Hernquist 1996).
Given the favorable scenario in which the red sequence is built up by ``wet mergers'' of late-type galaxies 
(e.g. Bell et al. 2004, Faber et al. 2007), one may expect a concentration 
of merging galaxies in the green valley, the area connecting the blue cloud and the red sequence 
in the color-magnitude diagram.  
The low fraction (6\%) of our ULIRGs in the green valley, however, suggests that most of these ULIRGs 
are not in the transition phase from the blue cloud to the red sequence. 
Instead, the majority (87\%) of these ULIRGs have typical blue-cloud colors, 
suggesting that they are still undergoing rapid 
star formation, and the significant quenching of star formation has yet to start. 

AGNs may play a crucial role in quenching star formation and 
accelerate the transition of blue-cloud galaxies to the red sequence 
(e.g. Hopkins et al. 2006; Croton et al. 2006). Indeed, some studies have found an excess of X-ray selected AGNs 
in the green valley (e.g. Georgakakis et al. 2008; Silverman et al. 2008; Hickox et al. 2009; Schawinski et al. 2009). None of the AGN ULIRGs in our sample, however, is located in the green valley, and only 
two ($17\%$) are redder than the blue limit of the red sequence. 
The remaining 10 ($83\%$) unsaturated sources are bluer than the red limit of the blue cloud.

Hickox et al. (2009) suggested an evolutionary sequence for massive galaxies with halo masses between 
$10^{12}\sim10^{13} \rm M_{\odot}$, in which a high accretion rate optical- and infrared-bright SMG/ULIRG/quasar phase
evolves quickly ($\lesssim 10^8\rm yr$), and is followed by a slightly lower accretion rate ``green-valley'' phase 
(lasting $\sim$ 1 Gyr) when the objects are detected as X-ray AGNs. 
This phase is followed by an intermittent radio AGN phase in which the galaxies reside in the red sequence
with lowest accretion rates. In the context of the Hickox et al. study, the majority of our AGN ULIRGs are still in the early 
optical-bright phase and have yet to evolve to the green valley.

\subsection{Color and Magnitude of AGN Host Galaxies}
\label{agn}

In the optical bands the AGN ULIRGs are among the most luminous in the ULIRG sample, and most have
blue colors (see \S\ref{cmr}). AGNs are point-like sources residing in the nucleus region and 
typically have blue optical colors.
Thus one may expect the presence of an AGN would affect the color and magnitude of the host galaxy 
(e.g. Hickox et al. 2009). However our AGN ULIRGs are low-luminosity Seyfert galaxies and 
the AGN emission at optical wavelengths may be highly attenuated. 
In order to test whether their high luminosity and blue color is due to 
a central AGN we estimate the AGN color and luminosity contribution by a simple model fitting.

\subsubsection{Methodology}

Because the resolution of the images 
is limited to only $\sim 1.3\arcsec$ (FWHM, a physical size of $\sim 2.4$ kpc at $z\sim0.1$, 
much larger than the physical size of the nuclear region, 
we do not try to decompose precisely the central AGN and the host galaxy light distribution from each image. 
Instead we subtract a scaled PSF from the nucleus of each galaxy and measure the flux of the scaled PSF and compare 
it with the flux of the host galaxy. As there are few unsaturated field stars on each image, 
we adopt the PSFs from the SDSS archive. We use two different subtraction methods and obtain 
a range of the flux for each subtracted scaled PSF and the host. 
In the maximum subtraction method, the maximally scaled PSF is subtracted from the galaxy image 
such that the residual is minimized but does not have negative pixels within a PSF FWHM size aperture 
around the center. By using this method we obtain 
an upper limit to the point source contribution and a lower limit of the host flux. 
In the second approach (smooth host subtraction method),
the PSF is scaled and subtracted such that the host has a smooth profile at the center
of the subtraction. The smooth host subtraction is based on the assumption that the galaxy is composed of a central point source
and a smooth host galaxy (e.g. McLeod \& McLeod 2001). 
We choose a 3 by 3 pixels area centered on the subtraction center and determine the host to be smooth 
if the standard deviation of these nine pixels is less than the average background rms noise on the same image.
For objects with multiple nuclei we subtract the scaled PSF from only the brightest nucleus. 
The exceptions are FSC13451+1232 (in $g$ and $r$ band) and FSC15001+1433 (in $g$ band), 
where the two nuclei are less than 4$\arcsec$ apart with flux difference less than 20\%. 
In these cases we subtracted scaled PSFs from both nuclei. We note that FSC12265+0219 and FSC12540+5708 
are saturated in both $g-$ and $r-$band images and are excluded from our analysis.

The flux of the subtracted point source is sensitive to its center position at the sub-pixel level. 
To find the center we first fit a two-dimensional gaussian profile for each source, and if the fitted FWHM is less than 30\% 
larger than the FWHM of the PSF, we adopt the fitted center as the subtraction center. 
Otherwise we adopt the centroid position as the subtraction center. 
The centroid is defined as the location where the spatial derivative of the 
pixel intensity becomes zero. 
The photometric uncertainty of the subtracted point source is calculated assuming it is dominated by the random noise within the central 
FWHM size aperture. The mean photometric uncertainties at $g$ and $r$ band are $\sim 0.05\rm\ mag$ and $\sim 0.02\rm\ mag$ for the 
maximum subtraction method, and $\sim 0.05\rm\ mag$ and $\sim 0.04\rm\ mag$ for the smooth host subtraction method, respectively.

\subsubsection{AGN Host Galaxies on the Color-magnitude Diagram}

On average the subtracted point sources contribute only a small amount to the total luminosity of
the AGN ULIRGs in both $g$ and $r$ band. This suggests that most of their optical luminosity
is associated with the stellar hosts. 
The maximum, median, and mean ratio of the subtracted point source to the 
total flux, using the maximum subtraction method, are 67\%, 28\%, 31\% in $g$ band, and 74\%, 27\%, and 31\% in $r$ band,
respectively. Using the smooth host subtraction method, these values are 61\%, 11\%, and 21\% in $g$ band, 74\%, 8\%, 
and 21\% in $r$ band, respectively. 
The maximum, median, and mean magnitude difference between the hosts and the un-subtracted 
sources are 1.19, 0.36, and 0.45 (1.48, 0.34, and 0.47) in $g$ band ($r$ band), using the maximum subtraction method, and
the corresponding values are 1.02, 0.12, and 0.31 (1.03, 0.09, and 0.31) in $g$ band ($r$ band), 
respectively, using the smooth host subtraction method.

We overlay the $^{0.1}g-^{0.1}r$ vs. $M_{^{0.1}r}$ color-magnitude diagram for the 12 AGN ULIRGs (filled circles) 
and their AGN-subtracted hosts (open circles) on the SDSS color-magnitude contours in Fig. \ref{agnhost}. 
Some hosts are much bluer after the subtraction (e.g. FSC11119+3257, bluer by 0.42 using both methods), some are
much redder (e.g. FSC01572+0009, redder by 0.33 using the maximum subtraction method). However,
the median color difference between the host and the original source is only 0.007 and 0.005 mag, 
using the maximum subtraction method and the smooth host method, respectively. 
Even after the maximum subtraction, only one (FSC11119+3257) source moved closer to the green valley, and the remaining
11 sources are located close to their original positions in the color-magnitude diagram and
barely overlapped with the SDSS comparison galaxies. 
This distinction is clearer using the smooth host subtraction method, 
after which the hosts are much closer to their un-subtracted counterparts. 
In either case most hosts (10 out of 12) remain luminous and blue in the 
color-magnitude diagram. With or without the point source subtraction, these AGN host ULIRGs do not appear in the green valley, 
the hypothesized transitional stage between the blue cloud and the red sequence that was associated
with X-ray and IR-selected AGNs (e.g. Hickox et al. 2009).
The blue colors of our AGN host galaxies are broadly consistent with other studies of optically-selected AGNs 
(e.g. Kauffmann et al. 2003; Jahnke et al. 2004; Silverman et al. 2008). 
Kaviraj (2008) studied a local LIRG ($L>10^{11}L_{\odot}$) sample and also found that virtually all LIRGs 
are in the blue cloud and the AGN has no impact on the star formation in their host galaxies.
In a broad agreement with out findings, Cowie \& Barger (2008) also found that the bulk of 
extinction-corrected 24$\mu$m selected sources in the GOODS-North field lie in the blue cloud.

Our study is limited by the small sample size, and the point source decomposition is limited by the spatial resolution of SDSS. 
A more complete sample observed with HST or the next generation James Webb Space Telescope ($JWST$; Gardner et al. 2006 ) 
is necessary in order to fully characterize the optical properties of the host galaxies in AGN ULIRGs.

\subsection{Color-magnitude Relation: ULIRGs vs Low-$z$ QSOs and Low-$z$ Type 2 Quasars}

Motivated by the high infrared and bolometric luminosity, it has been proposed that 
ULIRGs are the early stage of dusty quasars, and a ULIRG-QSO evolutionary scenario has been suggested 
(Sanders et al. 1988). Although the sources of the bolometric luminosity for ULIRGs are nearly completely 
obscured by dust in the visible bands, a comparison of their host properties with those of QSOs should  
provide an important test for the proposed evolutionary scenario. For example, if the ULIRGs and QSOs 
are identical except for a geometrical difference such as the viewing angle and dust geometry, 
then the optical properties of their hosts should be largely identical. If a purely temporal
evolution and an AGN feedback process removing gas and dust are the key difference 
(e.g. Narayanan et al. 2009), then a significant overlap with a systematic shift is expected in the 
host color-magnitude relation.

The QSO sample we selected is composed of 1070 objects in the SDSS DR5 quasar catalog 
(Schneider et al. 2007) within the same redshift range ($0.018 < z < 0.265$) as the ULIRGs. 
The SDSS QSOs are type 1 quasars, selected based on their broadband colors, with $M_i<-22$ mag and
spectroscopically confirmed to have broad lines (FWHM greater than 1000 $\rm kms^{-1}$). 
We also selected a type 2 quasar sample for comparison, which consists of 402 objects in the SDSS 
type 2 quasar catalog (Reyes et al. 2008) within the same redshift range, 
selected based on their optical emission lines.  
Systematic photometric difference should be minimal among the SDSS QSOs, type 
2  quasars, and the ULIRGs due to the shared origin of the data. For SDSS 
QSOs and type 2 quasars, we apply the k-correction and extinction correction 
using the same procedure as for the ULIRGs to obtain the absolute magnitudes. 
We plot in Fig. \ref{qso} the $^{0.1}g-^{0.1}r$ colors vs $M_{^{0.1}r}$ of the QSOs and the type 2 
quasars with dots and contours, on top of the contours of the SDSS field 
galaxies and the ULIRG symbols.

Strikingly, the distribution of the type 2 quasars peaks in the green valley where the X-ray 
and IR-selected AGNs are found (e.g. Hickox et al. 2009). In comparison, the SDSS QSOs only 
overlap slightly with the blue cloud and extend toward very blue colors. The difference in  
their distributions in the color-magnitude diagram is at least partly due  
to selection effects. The SDSS QSOs are selected unobscured type 1
quasars where the blue continuum and high luminosity of the central  
AGNs are visible, while the type 2 quasars are selected obscured AGNs
and thus the dust and gas block the view to the central blue and  
luminous AGNs. Studying whether these green type 2 quasars represent 
galaxies in a transitional stage from the blue cloud to the red sequence by 
AGN quenching, or their green colors are simply color combinations of redden  
obscured AGNs and the host galaxies, is beyond the scope of this paper.

The ULIRGs are distributed very differently from the SDSS QSOs and  
the type 2 quasars in the color-magnitude diagram. The mean absolute
magnitude of the ULIRGs ($\langle M_{^{0.1}r}\rangle=-21.4$) is very close to that of the SDSS  
QSOs ($\langle M_{^{0.1}r}\rangle=-21.3$), but their mean color ($\langle ^{0.1}g-^{0.1}r\rangle =0.57$) is 0.35 mag
redder than that of the SDSS QSOs (0.22). The SDSS QSOs occupy a  
narrow range in $M_{^{0.1}r}$, overlap only slightly with the ULIRGs in the color magnitude diagram, 
and extend farther to bluer colors than the ULIRGs. In comparison, the $\langle M_{^{0.1}r}\rangle$ of  
ULIRGs is 1.1 mag more luminous than that of the type 2 quasar  
(-20.3), and their $\langle ^{0.1}g-^{0.1}r\rangle$ is 0.14 mag bluer (0.57 compared to 0.74). 
The ULIRGs overlap with the type 2 quasars only at faint magnitudes
($M_{^{0.1}r}>-21$), and all but one AGN ULIRGs are more luminous than 90\% of  
the type 2 quasars. We also implement two-dimensional Kolmogorov-Smirnov tests 
(KS test, Peacock et al. 1983; Fasano \& Franceschini  
1987) between different samples in the color-magnitude two dimensional
parameter space, and the results are listed in Table \ref{KStable}. The test  
results indicate that neither the ULIRG sample nor the AGN-ULIRG  
subsample is drawn from the same underlying distribution with the SDSS QSO  
sample or with the type 2 quasar sample.

\subsection{Evolutionary Tracks on the Color-magnitude Diagram for ULIRGs}
\label{sedmodels}

We model the color-magnitude evolutionary tracks for the ULIRGs using the stellar synthesis model BC03 (Bruzual \& Charlot 2003) 
in order to gain some insight into the distribution and evolution scenarios of ULIRGs on the color-magnitude diagram.
Rather than providing a full solution to the SEDs with best-fit parameters, the primary goal of our models is to demonstrate 
qualitatively the possible evolutionary paths these galaxies may follow on the color-magnitude diagram 
and to establish the associated timescales. 

We start with the simplest cases in which the star formation 
histories are: 1) a passively evolved single stellar population burst (SSP), and 2) a constant star formation rate (constant). 
Fig. \ref{evolvtrack} shows the evolutionary tracks for these two models with a solid line (SSP) and a dashed line (constant).
Both tracks are k-corrected to $z=0.1$ in the same manner as the ULIRGs. The SSP track is normalized
so that it passes through the median color ($^{0.1}g-^{0.1}r = 0.58$) at the median absolute magnitude of the ULIRGs
($M_{^{0.1}r} = -21.43$, or $L_{r}=2.5L_{r,*}$). The constant track is normalized to reach 2.5$L_{*}$ at $t=20$Gyr.
We use solar metallicity in our models for simplicity and ignore intrinsic extinction
as we are primarily concerned with the unobscured stellar population.
Three symbols are plotted on each track to mark three ages, 1 Gyr, 2 Gyr, and 5 Gyr.
In the SSP model, we find that the stellar populations of ULIRGs have ages spanning from 
200 Myr to 8 Gyr based on their optical colors, with a median age of $\sim 1$ Gyr. 
One exception is the extremely red source FSC11119+3257 whose color is too red to be reproduced
by the SSP model. 
Our modeling suggests that a galaxy with a median age takes 1.5 Gyr to reach the red sequence.
In the constant star formation case, the reddest color of the stellar population (at $t=20$Gyr) is $^{0.1}g-^{0.1}r = 0.43$,
which is bluer than 80\% of the ULIRGs. This implies that in the constant star formation case 
the star formation has to be quenched in order to move the ULIRGs to the red sequence. 

Next we construct a model with a more complex star formation history
and apply different metallicity and dust treatments to try to understand their effects on the color and magnitude. 
Our model assumes that: (a) a ULIRG is a merger system consisting of two equal-mass late type galaxies
and that the merger triggers a starburst; (b) the star formation history consists of an SSP component associated 
with each progenitor and an exponentially decaying starburst triggered by the merger; (c) the SSP components are identical 
with the same age and the starburst is triggered when the SSP components evolve to the median 
color and magnitude of the ULIRGs; (d) the starburst forms as much as 10\% of its progenitors' stellar mass
with an e-folding time of $10^7$ yr. 
The timescale of $10^{7}$ yr is close to the dynamical timescale of ULIRGs such as Arp 220 ($6\times10^6$ yr;
Mauersberger et al. 1996) and 10\% is a fiducial fraction of mass formed during a merger that is 
consistent with the results in the numerical simulations of disk mergers (e.g. Mihos \& Hernquist 1996).  
Given that the progenitors are two $L_*$ galaxies with $M_* \sim 10^{11}\rm M_{\odot}$, 
the calculated star formation rate starts with a maximum value of $\sim2000\rm M_{\odot}yr^{-1}$
and decreases to several hundred $\rm M_{\odot}yr^{-1}$ after an e-folding time. 
Although these star formation rates seem extremely high, they are required by the observed high FIR luminosities.  
We calculate the star formation rate of the ULIRGs from their FIR luminosities using the 
FIR-SFR relation (Kennicutt 1998). The star formation rates range from 170 $\rm M_{\odot}yr^{-1}$ to 
1370 $\rm M_{\odot}yr^{-1}$, with a median of 270 $\rm M_{\odot}yr^{-1}$, which are consistent with our model predicted values. 
  
We examine three different metallicities, $Z=0.008$, $Z=0.02$, and $Z=0.05$, 
corresponding to a sub-solar, solar, and super-solar metallicity, respectively, 
and construct three base models using this 
star formation history. 
We assume that the starburst has the same metallicity as the passively evolved progenitor populations.
We set up a grid of different dust extinctions following Charlot \& Fall (2000) for each metallicity, 
but apply the dust extinction only to the starburst population. 
There are two parameters in the dust treatment method: the total effective V-band optical depth $\tau_{V}$ 
that affects stars younger than $10^{7}$ yr, 
and the fraction $\mu$ of extinction that comes from the diffuse ISM and thus affects stars of all ages. 
We choose $\tau_{V}$ between 1 to 10 with a step size of 1.
Bruzual \& Charlot (2003) noted that the average value of $\mu$ is 0.3, and we adopt this value for our modeling.
Our choice of optical depth is conservative, based on the common concept 
that these galaxies are dusty starbursts and/or AGNs. However we reiterate that the ULIRGs on average are blue and bright, 
and in order to produce the blue colors the dust attenuation 
associated with the escaping optical emission has to be modest.
As discussed earlier, it is also possible that the geometry of the dust distribution in the system is patchy 
so that the high infrared luminosity originates from the most dust-attenuated regions, while the optical emission 
comes from relatively less attenuated regions, e.g. tidal structures at large distances. 

We plot nine modeled tracks along with the ULIRGs on the color-magnitude diagram in Fig. \ref{fig8}.
The tracks start from the time when the starburst is induced. 
The three rows of panels show sub-solar, solar, and super-solar metallicity, 
from top to bottom, respectively. For each metallicity there are three tracks with $\tau_V\ = 1,\ 5,\ 9$, 
from left to right, respectively. The tracks are k-corrected to $z=0.1$ and normalized to the median $M_{^{0.1}r}$ 
which is roughly 2.5 $L_*$. Symbols are plotted on the tracks to mark the age of 1, 2, and 5 Gyr after the starburst starts. 
These tracks clearly suggest that both metallicity and dust extinction play important 
roles in our understanding of the evolution of these systems.
Dust extinction mostly affects the blue optical colors of the systems younger than several hundred million years. 
For all metallicities, the stellar populations reach their bluest colors in less than 20 Myr, but
the bluest color a stellar population can reach becomes increasingly redder with increasing dust extinction.
The blue colors of many ULIRGs can be produced only with the lowest dust extinctions. However, there is almost no color difference 
between different dust extinctions when the populations are older than 1 Gyr and their $^{0.1}g-^{0.1}r$ colors 
are redder than 0.7, and any systems with the same metallicity will redden to the red sequence at roughly the same time. 
Metallicity affects the range of colors and the timescale of the color evolution of the stellar populations, 
and higher metallicity systems span wider ranges in the color space, and evolve more quickly. 
As shown in Fig. \ref{fig8}, the bluest ULIRGs can be reproduced only by the highest metallicity track.
After 20 Gyr, the super-solar track is 0.25 mag redder than the sub-solar case.
It takes a sub-solar population 2 Gyr to evolve to the red sequence, 1.5 Gyr for the solar population, and only 1 Gyr for the
super-solar one. 

In reality the star formation history is almost certainly more complicated than the simple models we discuss here. 
In the third model we use a simplified version of star formation history produced from numerical simulations of
gaseous disk mergers by Springel, Di Matteo, and Hernquist (2005). Springel et al. used GADGET-2 (Springel 2005) to trace  
star formation in a merger of two spiral galaxies and studied the effects of back hole feedback on the quenching of star formation.
The two spiral galaxies have equal dynamical masses of $3.85\times 10^{12}\rm M_{\odot}$, 
and form stars at an exponentially decaying rate before
the merger, with the pre-merger star formation rate $\sim 200\rm M_{\odot}yr^{-1} $. The merger-triggered starburst
happens at $t=1.5\rm Gyr$ and the peak star formation rate is $\sim 2000\rm M_{\odot}yr^{-1}$. 
In the black hole feedback scenario the star formation is quenched right after the starburst and the star formation rate 
decreases to less than 1 $\rm M_{\odot}yr^{-1}$ at $t=2.5\rm Gyr$, while in the no feedback scenario the star formation rate 
decays slowly and remains at several solar masses per year for several Gyr. The intrinsic extinction is ignored in this model.
We plot the evolutionary tracks in Fig. \ref{evolvtrack} with the dotted line representing the track without AGN feedback and the 
dashed line representing the one with AGN feedback. Three symbols are plotted on each track to mark the ages of 1, 2, and 5 Gyr. 
We find that some of our most luminous ULIRGs are located very close to these tracks in the color-magnitude diagram.
Both tracks evolve more slowly than the SSP track on the same plot because the starburst produces new young blue stars.
Between the two tracks, the one with AGN feedback clearly reddens more quickly than the one without AGN feedback: 
it takes 8.5 Gyr (or 7 Gyr after the starburst) for the galaxies without AGN feedback to evolve to the red sequence, 
which is 1.5 Gyr longer than the one with AGN feedback. The difference in the timescales
is consistent with the scenario in which AGN feedback quenches the star formation effectively, and consequently helps form 
the color bimodality of galaxies.

In summary, these simple models suggest possible evolutionary paths of the ULIRGs on the color-magnitude diagram. We conclude 
that constant star forming galaxies cannot evolve to the green valley without the star formation quenched. Galaxies 
with lower star formation activity evolve to the red sequence more quickly and spend shorter time in the green valley, 
as seen in the SSP model and the numerical simulation model. Dust affects mostly the blue colors of the stellar population,
and has almost no influences on the red colors and the timescale of evolution to the red sequence. Stellar population 
with higher metallicity explores a wider range in the color space, and evolves more quickly to the red sequence. 

The color evolution of the ULIRGs may be presented by any one of the models with different masses and dust extinctions. 
For example, the evolutionary track of the most luminous ULIRGs in our sample may be presented by either the numerical simulation
models with slightly different mass, or by the SSP track in Fig. \ref{evolvtrack} with ten times more mass; the evolutionary 
track of a ULIRG with the median color and magnitude may be presented by the numerical simulation models with 90\% of the 
optical emission being attenuated. It is also possible that the color evolution of a galaxy is the result of composite 
star formation histories, so that the galaxy may leave the blue cloud and enter the red sequence for more than once. 
We will not discuss these more complicated scenarios because they are out of the scope of this paper.

\section{MORPHOLOGY}

\label{morphology}
\subsection{$Gini$ and $M_{20}$ Calculations}
\label{morph.calc}
We use the code developed by Lotz et al. (LPM04) to calculate the morphology coefficients $G$ and $M_{20}$
for both the 54 ULIRGs and the comparison sample, galaxies appearing on the same SDSS images
of the ULIRGs.  The Gini coefficient, $G$, is defined as
\begin{equation}
G = \frac{1}{\mid{\bar{X}}\mid n(n-1)}\sum_{i}^{n}(2i-n-1)\mid X_i\mid
\end{equation}
where $n$ is the number of pixels and $X_i$'s are the sorted pixel intensities in increasing order. The total second-order moment 
$M_{tot}$ of the pixels assigned to the galaxy is defined as
\begin{equation}
M_{tot} = \sum_{i}^{n}f_i[(x_i-x_c)^2+(y_i-y_c)^2],
\end{equation}
where $f_i$ is the flux in each pixel in decreasing order, $x_i$ and $y_i$ are the coordinates of the pixel, and $x_c$, $y_c$ is the center of the 
galaxy, computed by minimizing $M_{tot}$. $M_{20}$ is the normalized second-order moment of the brightest 20 percent of the galaxy's 
flux, such that
\begin{equation}
M_{20} = {\rm log}_{10}(\frac{\sum_i M_i}{M_{tot}})\           {\rm while}\            \sum_i f_i < 0.2f_{tot},
\end{equation}
where $f_{tot}$ is the total flux of the galaxy.

As noted by LPM04 and Lotz et al. (2008b), the quantity $G$ is very sensitive to the ratio of low surface brightness pixels 
to high surface brightness pixels, so a well-defined segmentation map is essential to measure $G$ and $M_{20}$ accurately. 
Also, as pointed out by Lisker (2008), measurement uncertainties of $G$ are minimized when 
using the Petrosian radius as the aperture size.
We thus adopt the same approach as LPM04 and assign the pixels brighter than the surface brightness at the Petrosian radius 
(semi-major axis as measured in an elliptical aperture) to the galaxy.
The measured coefficients are not robust for a source 
with the signal-to-noise ratio per pixel within the Petrosian radius ($\langle S/N \rangle $) 
less than 2.5, or with a Petrosian 
radius ($r_P$) less than twice the size of FWHM of the PSF ($\sim1.3\arcsec$ in SDSS) (LPM04 and Lotz et al. 2008b), 
where the signal-to-noise ratio per pixel is defined as the ratio of the mean flux and the mean rms noise within an aperture.
We therefore exclude all sources with $\langle S/N\rangle <2.5$, and/or $r_P<2.6\arcsec$. 
The $u-$band images are noisy in general: almost 60\% of the ULIRGs have $\langle $S/N$\rangle $ less than 2.5, and thus we do not 
include the $u-$band results further in our analysis. We also exclude possible candidates for stars in the comparison sample by rejecting 
sources with the 
$SExtractor$ star/galaxy separation parameter CLASS\_STAR greater than 0.9. The final selection yields the measured morphology parameters 
for 50, 50, 51, and 42 ULIRGs, and 1215, 2059, 2019, and 484 comparison galaxies, in the $g$, $r$, $i$, and $z$ band, respectively. 

\subsection{Morphology of the ULIRGs}

In Fig. \ref{gm20} we plot the $g-$, $r-$, $i-$, and $z-$band $G$ against $M_{20}$ of the comparison galaxies 
using contours and dots, where the contours are 10 equally spaced levels between 10\% and 90\% of the peak 
number density, and dots show galaxies located outside the lowest 10\% contour level. 
On top of them we overlay the morphology parameters of ULIRGs with circles (AGN ULIRGs) 
and stars (H{\sc ii}-region like/LINERs).
We also plot the empirical lines that divide the parameter space into four morphology regions: 
mergers, elliptical galaxies, irregular galaxies, 
and spiral galaxies, according to the classifications for $z\sim0$ galaxies described in LPM04 and Lotz et al. (2008a):
\begin{eqnarray}
\rm{Merger:} & G >  -0.115M_{20}+0.384 & \nonumber  \\ 
\rm{Elliptical:} & G < -0.115M_{20}+0.384 &\&\ G > 0.115M_{20}+0.769 \nonumber \\
\rm{Irregular:} & G > 0.115M_{20}+0.697 & \&\ G < 0.115M_{20}+0.769 \\ 
  & \&\ G < -0.115M_{20}+0.384 & \nonumber \\
\rm{Spiral:} &G < -0.115M_{20}+0.384 & \&\ G < 0.115M_{20}+0.697 \nonumber 
\end{eqnarray}

The distributions of the ULIRGs at all wavelengths are consistent in that they
form a similarly heterogeneous group in this parameter space. 
The centroids of the distribution 
in $G-M_{20}$ are [0.55,-1.42], [0.55,-1.43], [0.57,-1.47], [0.54,-1.45], in $g$, $r$, $i$, and $z$ band, respectively, 
all of which are located very close  ($\Delta G<\pm 0.02$ for any given $M_{20}$) to the separation line 
between merger and non-merger galaxies.  
There are only 21($42^{+18}_{-24}\%$), 21($42^{+20}_{-22}\%$), 25($49^{+19}_{-27}\%$), and 15($36^{+24}_{-26}\%$) ULIRGs 
located in the merger region, in $g$, $r$, $i$, and $z$ band, 
respectively. These are very low fractions compared to the $z\sim0$ results where the majority ($80\%$) of ULIRGs are located 
in the merger region (LPM04). The distributions of the ULIRGs in the parameter space vary slightly with wavelengths: 
the merger fraction is the highest ($49\%$) using the $i-$band derived parameters, and lowest ($36\%$) using the $z-$band derived ones.
Although the morphology parameters of the ULIRGs span a wide range in the parameter space, 
their $G-M_{20}$ distribution is clearly different from that of the comparison sample galaxies.
The distribution centroids of the comparison galaxies in the four bands are all located in the empirical region of spiral galaxies, 
and only 56($5^{+14}_{-4}\%$), 117($6^{+19}_{-5}\%$), 146($7^{+21}_{-6}\%$), and 26($5^{+21}_{5}\%$) galaxies are found in the empirical region of mergers, 
in $g$, $r$, $i$, and $z$ band, respectively.

The poor physical resolution of the SDSS images 
(700 pc pixel$^{-1}$ at $z\sim 0.1$) may lead to higher random and systematic uncertainties in $G$ and $M_{20}$ 
(see LPM04), and thus careful estimation 
of the uncertainties in $G$ and $M_{20}$ is necessary. 
Using image and noise simulations (described below in \S\ref{morph.discussion}), we estimate the uncertainties of 
morphology measurements ($\Delta G\lesssim0.05$ and $M_{20}\lesssim0.3-0.4$) 
in SDSS images and plot the largest uncertainties as error bars in Fig. \ref{gm20}. 
We find that a galaxy seen against higher background noise tends to move
to the left lower corner in the $G-M_{20}$ plot, and we plot a vector to illustrate the typical changes in 
$G$ and $M_{20}$ resulting from adding background noise until the source only barely satisfies the
signal-to-noise ratio and Petrosian radius criteria.
The vector shows that a source selected as a merger at high signal-to-noise ratio could move away from the 
merger region and stay in the late-type galaxy region if it is immersed in higher background, and consequently,
the merger fraction selected by $G$ and $M_{20}$ is likely to be a lower limit. 
Although less than half of the ULIRGs have empirical merger $G-M_{20}$ relations, the heterogeneous 
distributions are qualitatively consistent with what was found in the recent numerical simulations 
of merging galaxies of Lotz et al. (2008b). We will discuss these noise simulations further, together with the uncertainties 
in measuring the morphologies, in \S\ref{morph.discussion}.

AGN ULIRGs are slightly more concentrated in the merger region than non-AGN ULIRGs: there are 45\%, 50\%, 60\%, 50\% of 
AGN ULIRGs in the merger region, in $g$, $r$, $i$, and $z$ band, respectively, compared to 41\%, 40\%, 48\%, 30\%
for non-AGN ULIRGs. 
However the difference is not statistically significant given the small number of AGN ULIRGs (11 in $g$ band
and 10 in other bands based on the signal-to-noise ratio and size selection criteria). 
There is no significant difference between the mean $G$ ($\langle G\rangle$) and $M_{20}$ ($\langle M_{20}\rangle$) 
for the AGN ULIRGs and non-AGN ULIRGs: for AGN ULIRGs, $\langle G\rangle$ is 0.58, 0.57, 0.59, 0.58, 
and $\langle M_{20}\rangle$ is -1.48, -1.48, -1.36, -1.48, in $g$, $r$, $i$, and $z$ band, respectively; 
in comparison, $\langle G\rangle$ is 0.54, 0.54, 0.56, 0.53, and 
$\langle M_{20}\rangle$ is -1.42, -1.45, -1.49, -1.42 for non-AGN ULIRGs in the same bands.

\subsection{Are the ULIRGs Mergers in $G-M_{20}$ space?}
\label{morph.discussion}

\begin{figure}
\includegraphics[width=6in]{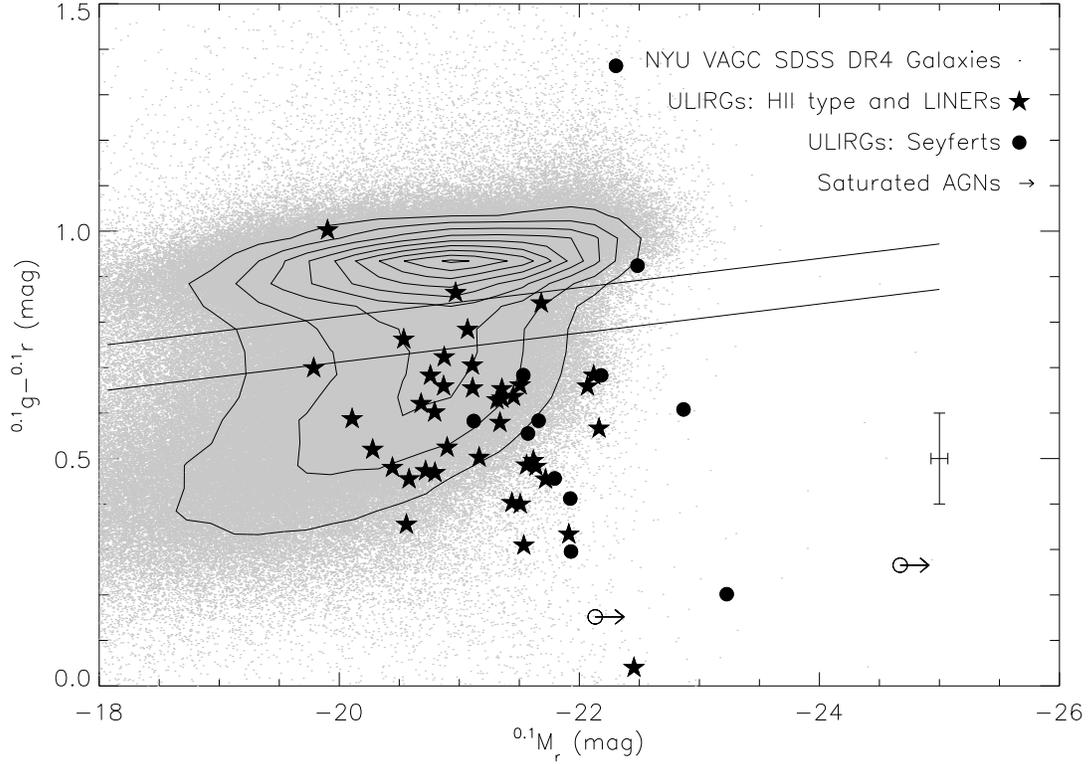}
\caption{Color-magnitude diagram derived from SDSS. Scattered gray dots represent NYU VAGC DR4 galaxies with known spectroscopic redshifts within $0.018<z<0.265$, 
the same redshift range of the ULIRGs. Contours represent 10 equally spaced levels between the minimum and maximum number 
density. Filled circles and stars represent the ULIRG sample, with circles representing AGNs and stars representing H{\sc ii}-like and LINERs,
respectively. The solid line is taken from Weinmann et al. (2006) as an empirical separation between the red sequence and the blue clouds. 
Source $FSC 12265+0219$ and $FSC 12540+5708$ are saturated in the $g$ and $r$ band and open circles and arrows are plotted 
to denote the lower limits. The error bar shows typical photometric uncertainties for ULIRGs only, 
and is dominated by aperture differences among different bands. 
See text for more detailed descriptions on the uncertainties. Typical photometric
uncertainties for the SDSS comparison sample are less than 0.03 mag in $g$ and $r-$band (York et al. 2000).
The plot shows that 24, or $\sim$43\% of the ULIRGs lie outside the 90\% normal galaxy contour,  $\sim$ 46\% have the typical blue-cloud
color, and only $\sim$7\% fall in the red sequence. Strikingly, only 3 ($\sim6\%$) ULIRGs are located in the green valley, and none of which
is an AGN ULIRGs. See text for details.}
\label{fig2}
\end{figure}

\begin{figure}
\begin{center}
\includegraphics[angle=0, width=18cm]{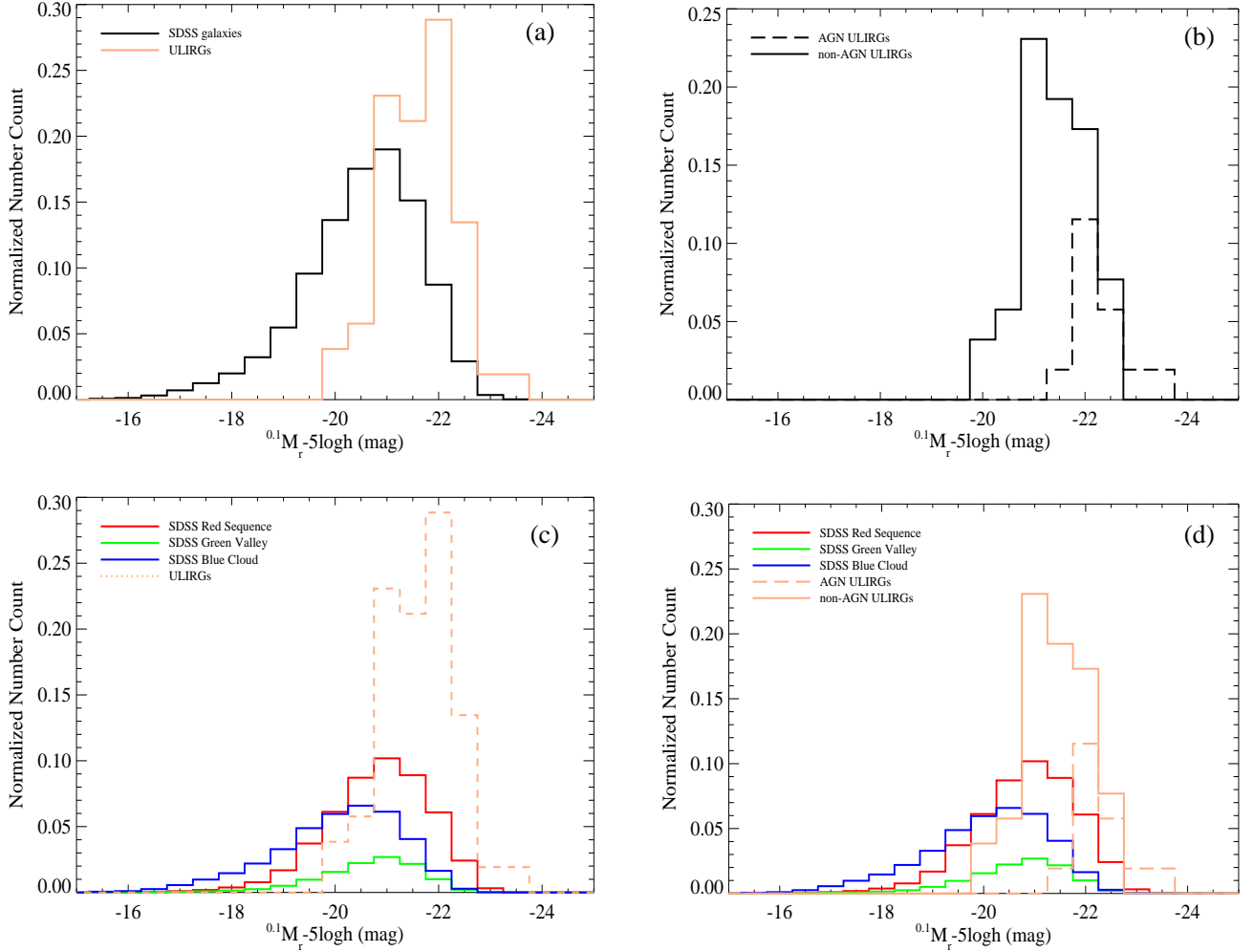}
\end{center}
\caption{K-corrected absolute magnitude ($M_{^{0.1}r}-5\rm log\it h$) distributions for $(a)$ the ULIRG sample (beige solid line) and 
the SDSS comparison sample galaxies (black solid line); $(b)$ the AGN ULIRGs (solid line)
and non-AGN ULIRGs (dashed line); $(c)$ the ULIRG sample (beige dashed line)
and the three sub-samples of the SDSS comparison sample (solid lines: the red sequence, red; the green valley, green; and the blue cloud, blue); 
$(d)$ the two subsamples of the ULIRG sample (AGN ULIRGs: beige solid line; non-AGN ULIRGs: beige dashed line) 
and the three sub-samples of the SDSS comparison sample (the same symbols as in $(c)$).
Note that there is very little overlap between the AGN ULIRGs and the green valley in magnitude distribution.} 
\label{maghist}
\end{figure}

\begin{figure}
\begin{center}
\includegraphics[angle=0, width=18cm]{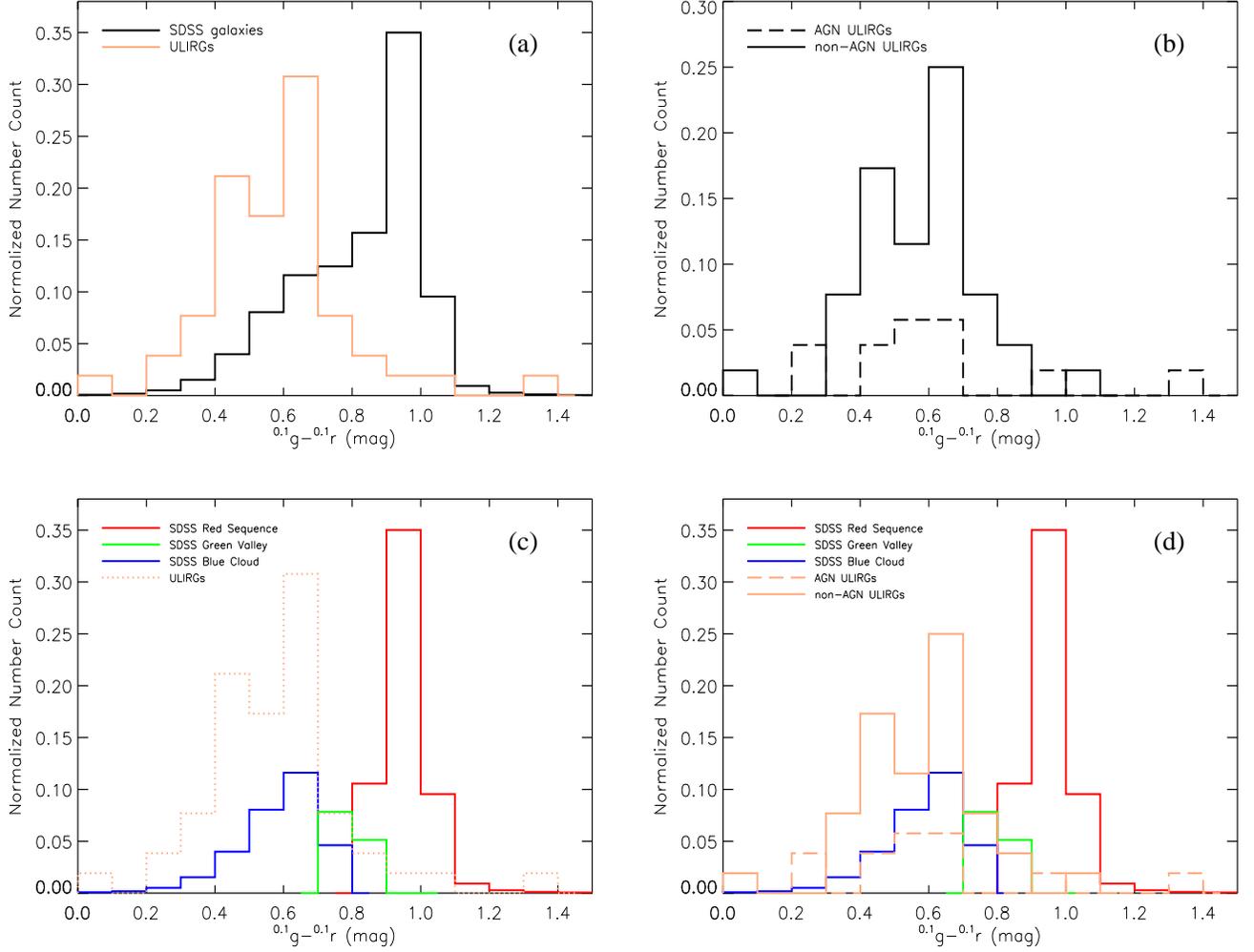}
\end{center}
\caption{$^{0.1}g-^{0.1}r$ color distributions for $(a)$ the ULIRG sample and the SDSS comparison sample galaxies within the same magnitude range 
($-19.8<M_{^{0.1}r}-5logh<-23.2$); $(b)$ the AGN ULIRGs and non-AGN ULIRGs; $(c)$ the ULIRG sample
and the three sub-samples of the SDSS comparison sample (the red sequence, green valley, and blue cloud) within the same magnitude range 
($-19.8<M_{^{0.1}r}-5logh<-23.2$); 
$(d)$ the two subsamples of the ULIRG sample and the three sub-samples of the SDSS comparison sample within the same magnitude range 
($-19.8<M_{^{0.1}r}-5logh<-23.2$). 
The symbols are the same as in Fig. \ref{maghist}. Note that there is little overlap between the AGN ULIRGs and the green valley in color distribution. }
\label{colorhist}
\end{figure}

\begin{figure}
\includegraphics[width=18cm]{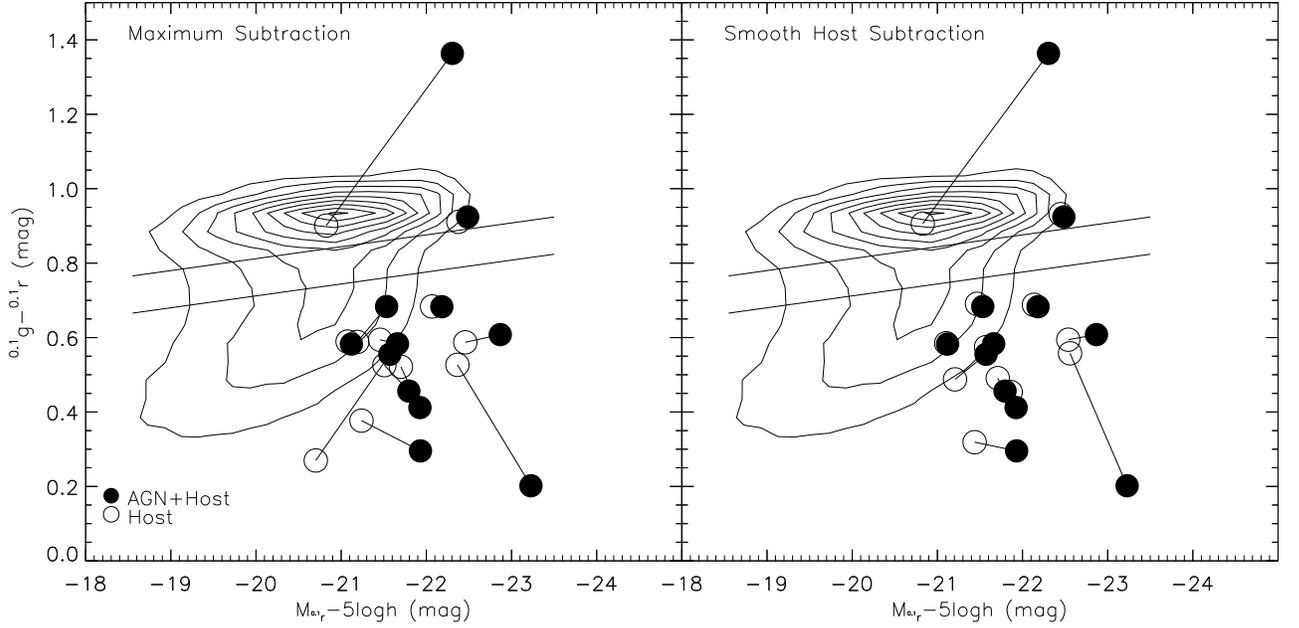}
\caption{Color-magnitude diagram for the 12 unsaturated AGN ULIRGs (filled circles) and the residuals after the point source subtraction
(empty circles), each pair connecting with a solid line, superimposed on the number density contours of the SDSS comparison sample galaxies.
The area between the two straight lines is the green valley. {\it left:} maximum subtraction method;
{\it right:} smooth host subtraction method. See \S\ref{agn} for details about the subtraction methods.
The subtracted point sources contribute on average a small fraction to the total flux, and the residuals are on average
only 0.45 and 0.47 mag fainter than their un-subtracted counterparts in $g$ and $r$ band, using the maximum subtraction method,
and 0.27 and 0.31 mag fainter in $g$ and $r$ band, using the smooth host method, respectively. Although some sources do appear to
be much redder or bluer after the point sources being subtracted, the average $^{0.1}g-^{0.1}r$ color is almost the same.
The residuals are barely overlapped with the SDSS comparison galaxies and we do not observe an concentration of AGN host galaxies
in the green valley.}
\label{agnhost}
\end{figure}

\begin{figure}
\includegraphics{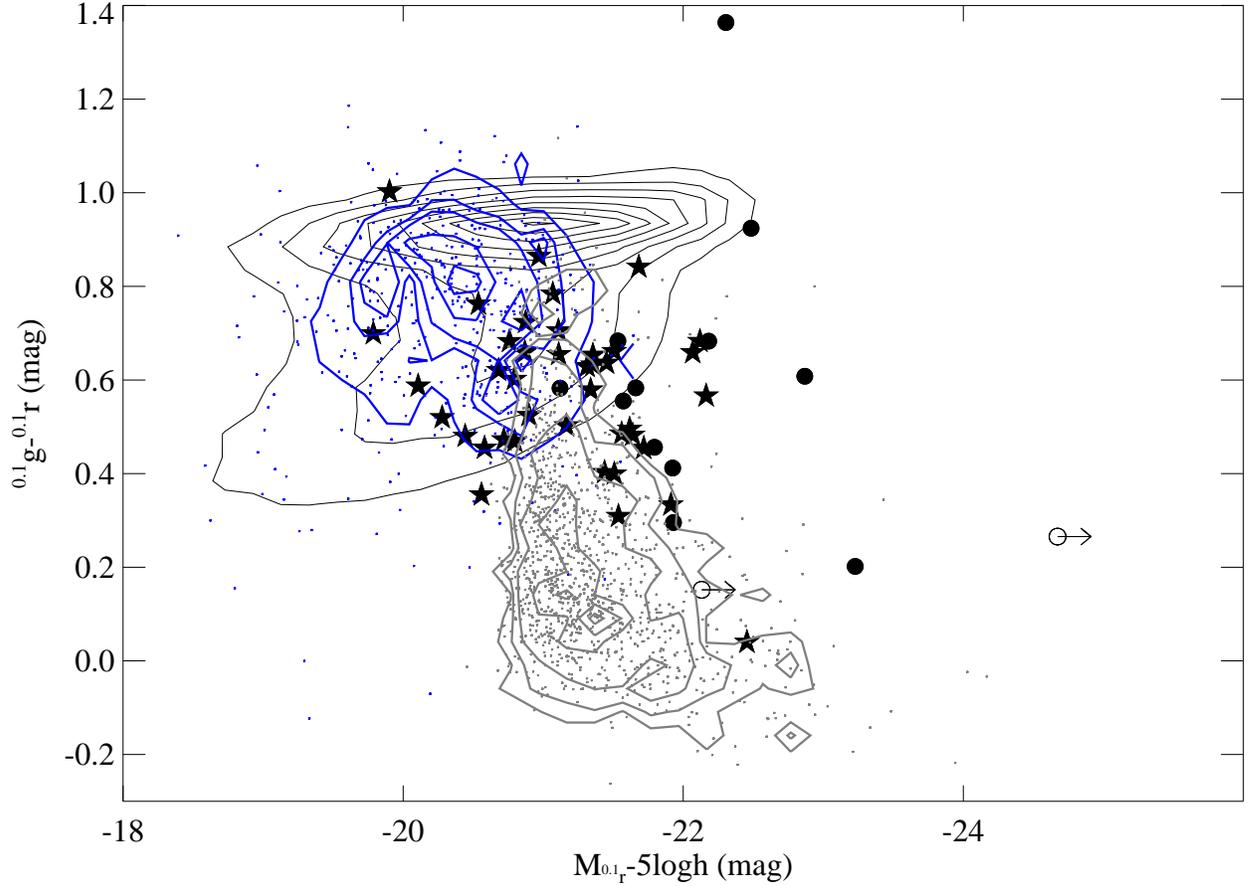}
\caption{Black contours and symbols represent the SDSS field galaxies and the ULIRGs, 
respectively. Grey contours and dots represent the SDSS QSOs within the same redshift 
range of the ULIRGs, and blue contours and dots represent the SDSS type 2 quasars within the same redshift range. The 
ULIRGs have comparable optical luminosity to the low-redshift QSOs, but the QSOs extend to much bluer
regions in the color-magnitude diagram. The ULIRGs are more luminous and redder than the SDSS type 2
quasars. Strikingly, the distribution of the SDSS type 2 quasars peaks at the green valley. }
\label{qso}
\end{figure}

\begin{figure}
\includegraphics{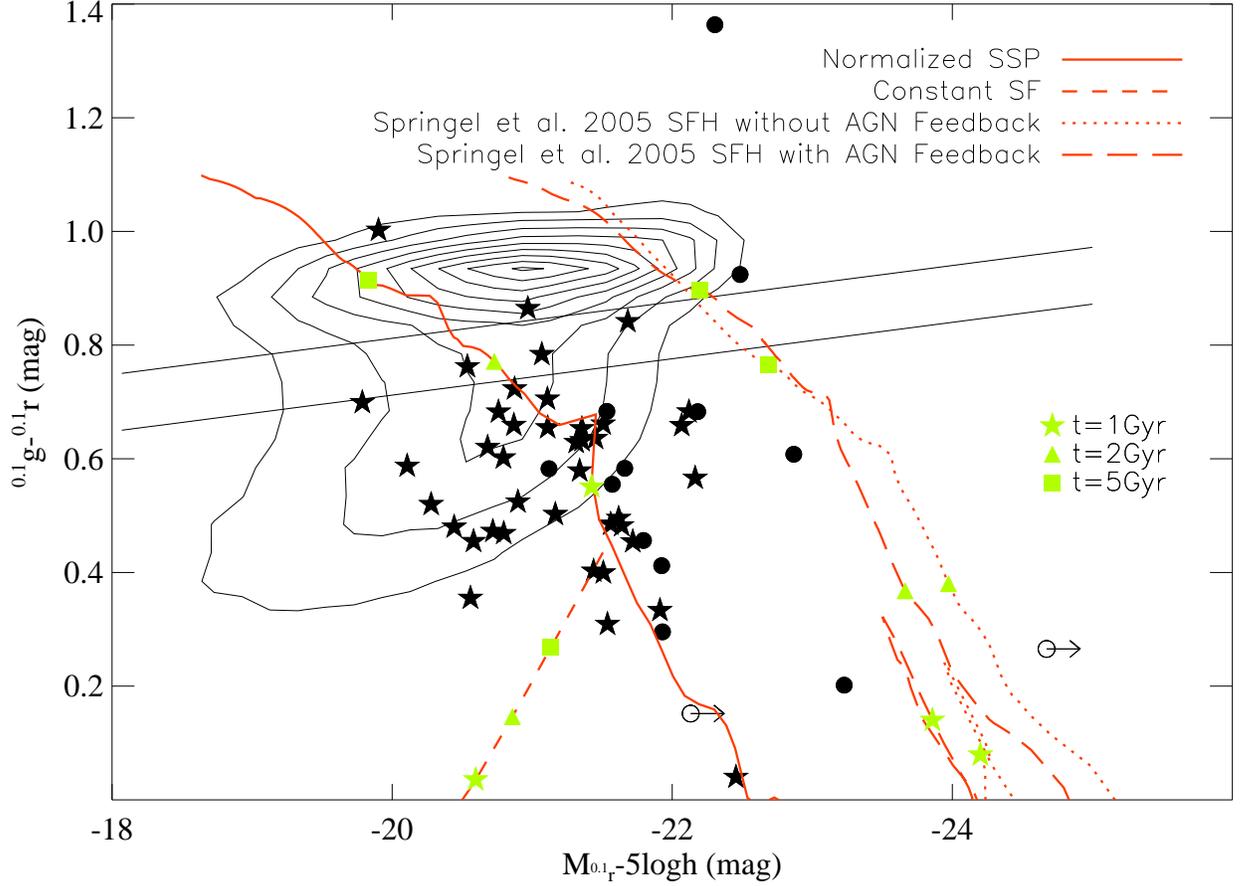}
\caption{Evolutionary tracks are superimposed on the color-magnitude diagram, where the
contours and symbols are the same as in Fig.\ref{fig2}. Different tracks represent different star formation histories:
normalized single stellar burst (SSP, solid line; see \S\ref{sedmodels} for details),
constant star formation (dashed line), star formation histories predicted by numerical simulations of the merger of
two spiral galaxies (Springel et al. 2005), each galaxy has a dynamical mass of $3.85\times10^{12}M_{\odot}$,
with (long dashed line) and without (dotted line) AGN activity. The color and magnitude at 1, 2, and 5 Gyr
are marked by stars, triangles, and squares, respectively, for each of the four tracks. The evolutionary tracks are calculated
using Bruzual \& Charlot (2003) with the Salpeter IMF and solar metallicity. }
\label{evolvtrack}
\end{figure}

\begin{figure}
\includegraphics{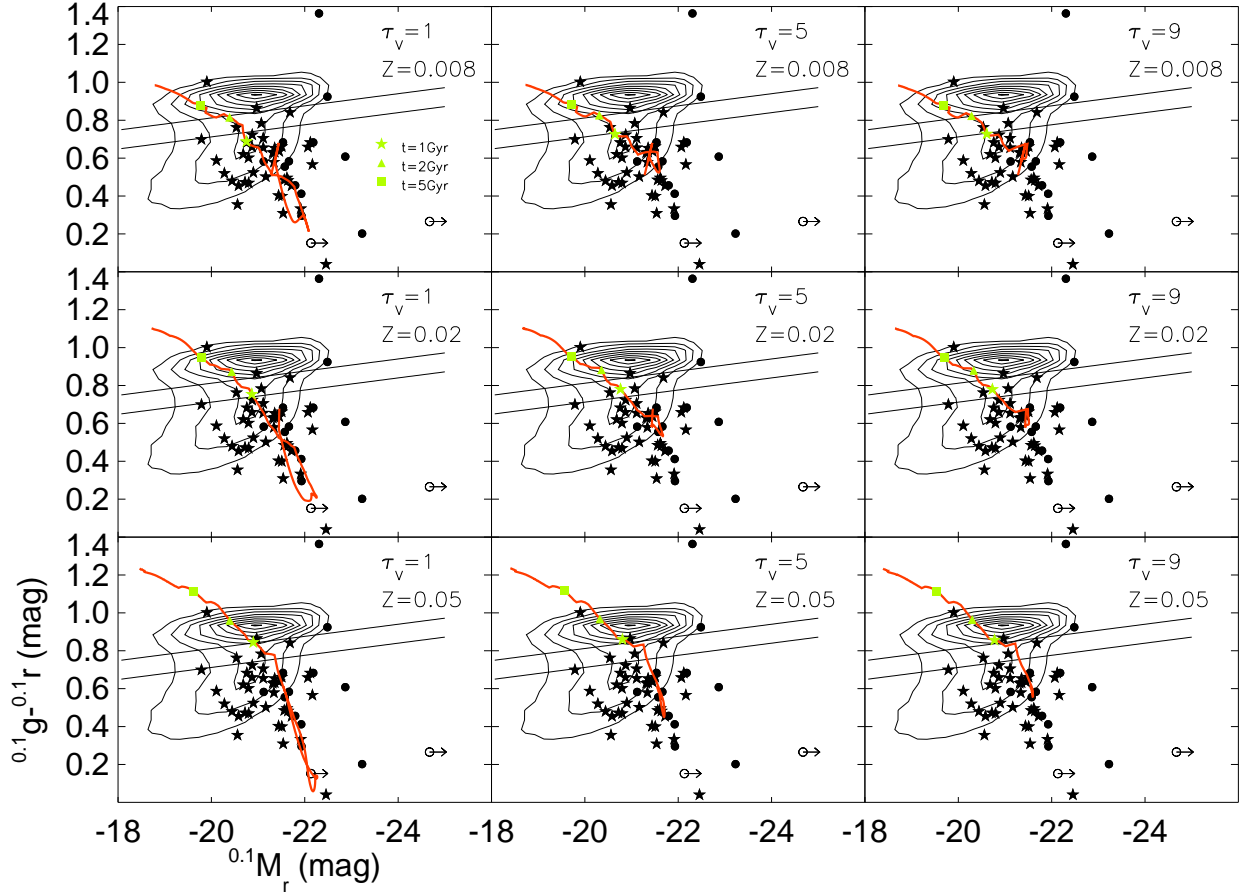}
\caption{Evolutionary tracks of a combination of a merger-triggered starburst and its passively evolved progenitors,
with different
metallicities and dust extinctions. See \S\ref{sedmodels} for more details. Three
symbols are plotted on each track to mark 1Gyr, 2 Gyr, and 5Gyr {\it after} the
starburst. We find that dust extinction mainly affects blue optical colors
at early ages of the stellar populations, and has little effect after 1 Gyr.
The metallicity affects both the color range of the stellar populations and
the evolutionary timescale. The stellar populations with the highest metallicity
(bottom panels) evolve to the red sequence twice as fast as the
lowest metallicity ones (upper panels).}
\label{fig8}
\end{figure}

\begin{figure}
\includegraphics[width=15cm]{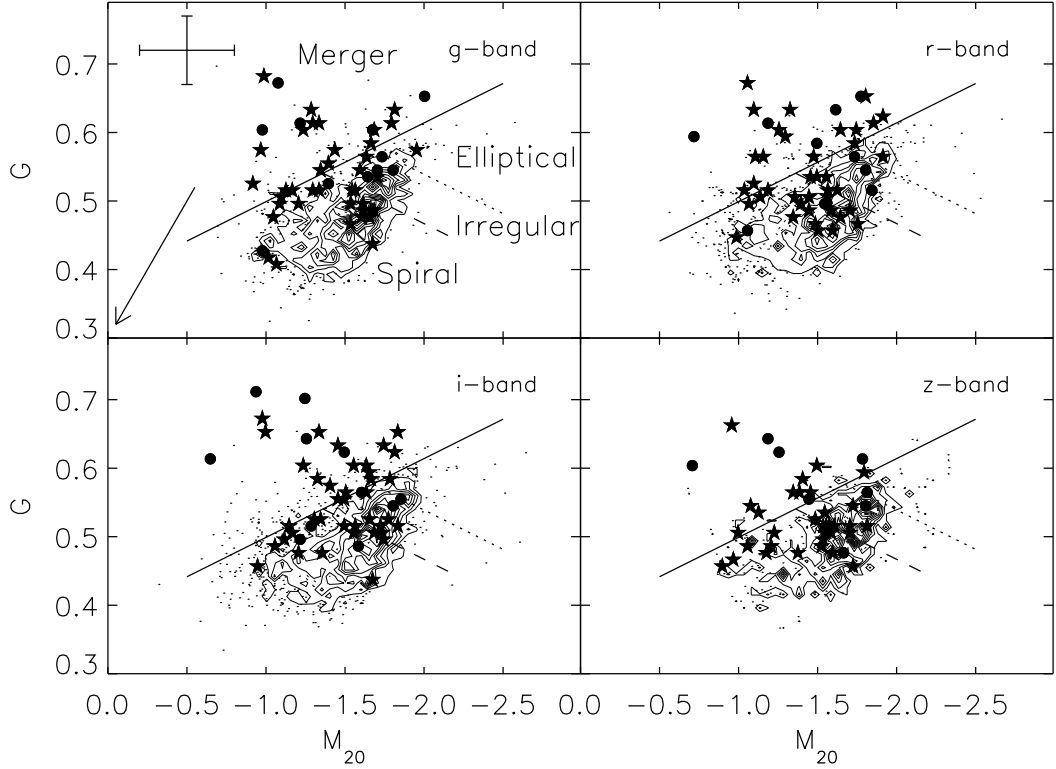}
\caption{$G-M_{20}$ of the ULIRG sample shown with circles (AGN ULIRGs) and stars (non-AGN ULIRGs), superimposed on the comparison sample 
(dots and contours) in $g$, $r$, $i$, and $z$ bands. The comparison sample consists of galaxies on the same images as the ULIRGs, 
excluding possible stellar objects (see text for detailed descriptions). Only sources with signal-to-noise ratio per pixel greater
than 2.5, and Petrosian radius greater than twice the size of the PSF FWHM ($\sim2.6\arcsec$) are plotted. 
The solid line, dashed line and dotted line divide the parameter space into four different regions 
(as marked in the $g$-band panel) according to LPM04: merger, elliptical, irregular and spiral. 
AGN ULIRGs are not more preferentially concentrated in the merger 
(non-merger) region than the non-AGN ULIRGs. Error bars plotted in the upper left panel show the maximum uncertainties 
in $G$ and $M_{20}$ estimated from our simulations.
The vector in the same panel indicates the upper limits of systematic changes in the measurements
of $G$ and $M_{20}$ given increasing background noise until the galaxy is no longer detected
(either $r_p < 2 \times FWHM$, or $\langle S/N\rangle < 2.5$). Please refer to \S\ref{morph.discussion} for a more detailed description.}

\label{gm20}
\end{figure}

Our visual inspection shows that almost all the ULIRGs are disturbed systems (Fig. \ref{fig2}), and this is consistent with 
the visual classifications by VKS02, who found that all but one of the 118 $IRAS$ 1Jy sample show visual merger features.
We perform the $G-M_{20}$ analysis and find that more disturbed galaxies do have higher $G$ and $M_{20}$ coefficients. 
This is illustrated in Fig. \ref{fig1} where 
more disturbed galaxies tend to occupy the upper left corner of the mosaic with higher values of both $G$ and $M_{20}$.
However our studies also show that ULIRGs are a heterogeneous group in $G-M_{20}$ space: only slightly less than half of the 
sources lie above the solid line in each panel of Fig. \ref{gm20}, the region where most local mergers and ULIRGs 
were found by LPM04. Interestingly, there is only one source located in the early type region, 
and all the remaining sources fall in the region where irregular and late-type galaxies are located.
This heterogeneous distribution seems to be inconsistent with the result of LPM04
that 80\% of local ULIRGs are located in the typical merger regions in the parameter space.
However, the spread in the parameter space we found is supported by numerical simulations. 
Lotz et al. (2008b) analyzed the morphological parameters in SDSS $g-$band images of mergers of equal mass gas-rich 
spirals by using the N-body/hydrodynamic simulation code GADGET (Springel, Yoshida, \& White 2001) 
and Monte Carlo radiative transfer code SUNRISE (Jonsson 2006; Jonsson et al. 2006), 
and find that the mergers are most disturbed in $G-M_{20}$ at the first pass, 
but they can have normal galaxies morphologies at other merger stages.
They also noted that two-thirds of the ULIRGs in LPM04 exhibit double or multiple nuclei, 
and therefore are more effectively selected by $G-M_{20}$.
The timescale is only 0.2 $-$ 0.6 Gyr for a merger to appear in the empirical merger region defined by LPM04, 
and the range in this time scale depends on dust, orbit parameters, and observing orientations. 
Thus at face value, our results imply that about half of our ULIRGs have been captured within this time window. 
The fact that the other
half of our ULIRGs do not appear in the empirical merger region does not mean that they are not mergers. Instead, they might be in  
other merger stages when their morphological parameters are consistent with those of normal galaxies. The heterogeneous picture 
of our ULIRGs thus may represent different evolutionary epochs of the ULIRGs.

\begin{figure}
\includegraphics{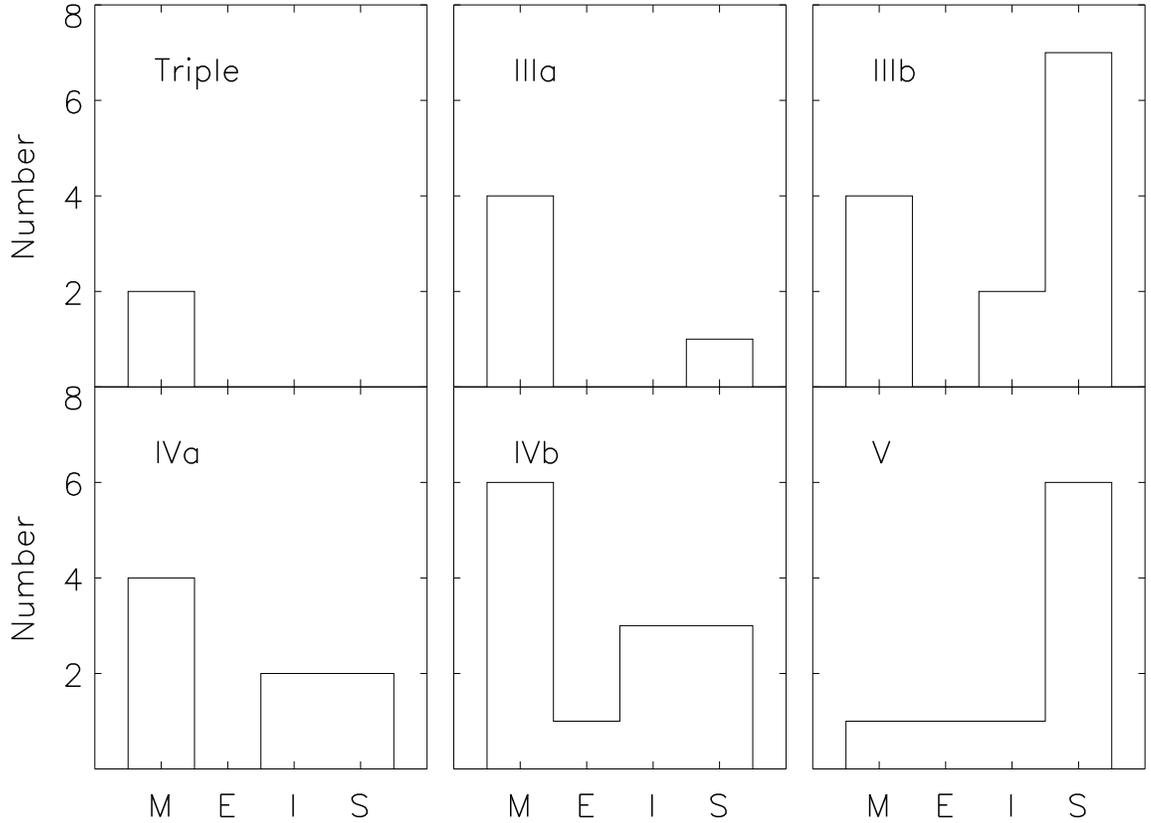}
\caption{Distribution of morphology classifications using $G-M_{20}$ at each merger stage defined in Veilleux et al. (2002) for the same sources.
The merging stages are: Tpl - triple mergers, IIIa - wide binary pre-mergers, IIIb - close binary pre-mergers, IVa - compact mergersm, IVb - diffuse
mergers, V - old mergers. Values in the horizontal axis stand for the four $G-M_{20}$ classifications: M (merger), E (elliptical), I (irregular), and S (Spiral). }
\label{morph.comp}
\end{figure}

In order to gain some insights into the $G-M_{20}$ morphology
we compare our $g$-band $G-M_{20}$ classification with the interaction
classification for the same sources made by Veilleux et al. (2002).
Veilleux et al. classified each object into one of the six sequential
merging stages according to its morphology. The comparison of two different
classifications is illustrated in Fig. \ref{morph.comp}. We find that all triple mergers by the
classification of Veilleux et al. are also classified as mergers by $G-M_{20}$.
Sources in their type IIIa group, i.e., wide binary pre-merger systems
with apparent separations greater than 10 kpc, are also very likely to be
recognized as $G-M_{20}$ mergers: $\sim 80\%$ of such sources are
recognized as mergers using $G-M_{20}$. Sources in their other groups
are less likely to be recognized by $G-M_{20}$ as mergers, and the fraction
of $G-M_{20}$ mergers appears to decrease toward later merging stages.
We find that $G-M_{20}$ mergers make up $\sim 31\%$ of the close binary pre mergers (type IIIb), 
$\sim50\%$ of the compact mergers (type IVa), $\sim 46\%$ of the
diffuse mergers (type IVb), and only $\sim 11\%$ of the old mergers (type V)
are $G-M_{20}$ mergers, respectively. Sources classified as close binary
pre-merger systems (type IIIb) and old mergers (type V) have very large
fractions to be recognized as spiral galaxies by $G-M_{20}$ ($\sim 54\%$ and $\sim 67\%$,
respectively). These comparison results are qualitatively consistent with
the simulation results by Lotz et al. (2008) in which the most disturbed $G-M_{20}$ morphology
happens during the first pass and maximum separation. However, since our measured morphology is limited by
relatively small number statistics and relatively large uncertainties, we do not conclude
strongly that the observed morphology can be well explained by the simulations.

We also find that uncertainties in measuring the morphologies contribute significantly to the distribution of the ULIRGs in the parameter space. 
LPM04 reported that the measurements of $G$ and $M_{20}$ are robust to within 10\% 
when the source has signal-to-noise ratio per pixel $\langle S/N\rangle $ greater than 2.5, 
the Petrosian radius larger than 2.5 times the PSF FWHM, and the physical resolution (parsec per pixel) higher than 
500 $\rm pc\ pixel^{-1}$. However when the physical resolution is lower the uncertainties of morphologies largely increase 
(Fig. 6 in their paper)
since small structures are washed out. At $z\sim 0.1-0.2$ the spatial resolution of the SDSS ULIRGs is roughly 0.7-1.3 $\rm kpc\ pixel^{-1}$
and the uncertainties need to be calibrated carefully. 
  
We perform simple simulations to estimate the uncertainties of $G$ and $M_{20}$ for the SDSS ULIRGs at different noise levels.
We use four model elliptical galaxies, four spiral galaxies selected from SDSS images, and three merging galaxies 
selected from our 
ULIRG sample. The four elliptical galaxies are modeled using de Vacoulers profiles and are added to real SDSS images to mimic the observational
conditions. The three spiral galaxies are selected with bright $g$-band magnitudes of 13.9, 15.2, and 16.4, respectively, much brighter than the 
median $g$-band magnitudes of the SDSS sample (18.0 mag). The three ULIRGs (FSC08572+3915, FSC14060+2919, FSC14121-0126)
are selected with bright $g$-band magnitudes and unmistakable merger morphology. The $g$-band magnitudes are 16.4, 16.7 and 17.8, respectively, 
all brighter than the median g-band magnitudes of the ULIRGs (17.8 mag).
For each image we generate random noise at a series of different levels, and at each level we generate 20 noise maps independently 
and add them to the original image.
Morphologies, Petrosian radius, and $\langle S/N\rangle$ are measured for each noise-added source and compared to the original measurements.
Although our simulation is limited by the sample size, $\sim90\%$ of the noise-added sources selected 
by the signal-to-noise and size criteria ($\langle S/N\rangle>2.5$ and $R_p>2.6\arcsec$) have $g-$band magnitudes between 16 and 19, 
and $g-$band Petrosian radii between 3 and 12 arcsec ($\sim$ 6 - 22  kpc at $z\sim0.1$), ranges similar to those of the ULIRGs.

We find that the measured Petrosian radius and the $\langle S/N\rangle$ do not always decrease monotonically with the amount of noise
added, and they are broadly distributed at low signal-to-noise levels. We also find that, rather than remaining fixed for a given source, 
the measured $G$ decreases and $M_{20}$ increases systematically with increasing noise. Therefore when seen against higher 
background noise, a galaxy tends to move to the lower left corner on the $G-M_{20}$ plot.
This effect will decrease the fraction of the mergers observed in the merger region.
The Gini coefficient $G$ decreases by as much as 0.2 with increasing background noise until the galaxy is no longer detected
(either $r_p < 2 \times FWHM$, or $\langle S/N\rangle < 2.5$), and $M_{20}$ increases
by as much as 0.5. The standard deviations of $G$ at every noise level are all less than 0.05,
and the spiral galaxies have the smallest dispersions. The standard deviations of $M_{20}$ are all less than 0.3-0.4, 
and the elliptical galaxies have the smallest dispersions. There is no systematic difference for the uncertainties between different bands. 
These uncertainties ($\Delta G=0.05$ and $\Delta M=0.5$) are shown in Fig. \ref{gm20} as error bars, and the systematic
trend with increasing noise in $G$ and $M_{20}$ is shown as an arrow.

The aperture within which the morphologies are measured is also an important source of 
uncertainty. As noted by Lisker (2008), $G$ value measured within larger apertures have systematically larger
values, and the measured uncertainties are minimized when the Petrosian radius is used as the aperture size. 
The reason behind is that more low surface brightness pixels are included within a larger aperture size, 
which steepens the intensity distribution of the pixels and consequently increases the value of $G$. 
Some of the low surface brightness features of the source, often at larger distances from the center, are mistakenly excluded 
when the aperture size is chosen too small, which flattens the pixel intensity distribution and consequently decreases the value 
of $G$. When assigning pixels to the sources we adopt the surface brightness at the Petrosian radius as the threshold, and
assign pixels brighter than the threshold to the source. Therefore, according to their study our measured morphologies should
have minimized uncertainties, because our apertures should be very close to the Petrosian radius and not significantly 
larger or smaller.

\subsection{Relations among Morphology, Optical Color, and FIR Luminosity}

The morphology, color-magnitude relation, and the infrared luminosity of ULIRGs present 
a rather complicated picture of these systems. The huge infrared luminosity in our ULIRGs suggests 
that their dominant powering sources should be obscured by dust. However, 
the majority of them are optically bright and blue. This further 
implies that dust is not uniformly distributed 
and that significant amount of the unobscured stellar light is seen directly.
Indeed, as shown in Fig. 1, there are obvious color gradients in the ULIRGs, with their tidal features at large 
distances appearing blue and their central regions appearing red. Lotz et al. (2008b) suggested
that merging systems are most disturbed in $G-M_{20}$ space and are located in the merger region in the $G-M_{20}$ plot
during the first pass when the tidal features are prominent. 
After the first pass they will move to the late-type and irregular region (see Fig. 5 in their paper). 
Thus we would expect the integrated colors of those disturbed systems 
selected by $G-M_{20}$ to be relatively bluer than the ones with normal $G-M_{20}$ relations, 
due to the blue colors of the tidal features. Recent hydrodynamic simulations of disc-disc 
mergers also suggested that the infrared luminosity of ULIRGs will peak during the final 
merger when enough metals and dust have been accumulated (e.g. Jonsson et al. 2006),
so one would also expect sources with higher infrared luminosity are preferentially found in the late-type and irregular region.

\begin{figure}
\includegraphics{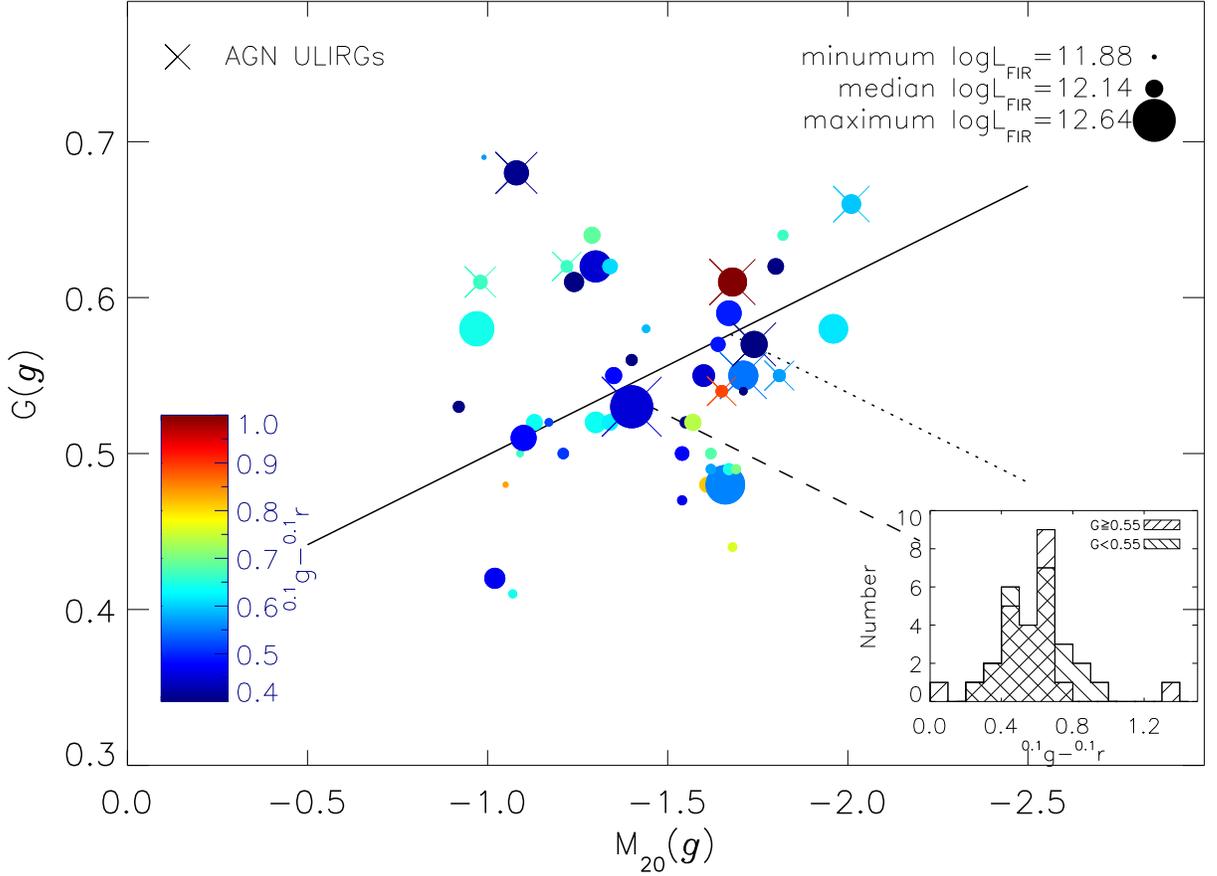}
\caption{$G$-$M_{20}$ measured in $g-$band. Different symbol sizes representing different far-infrared luminosities, linearly
in logarithm scale. Different symbol colors representing different $^{0.1}g-^{0.1}r$ colors. The small panel shows the 
distributions of more disturbed systems ($G\ge0.55$) and of less disturbed systems ($G<0.55$). Sources with redder optical colors
are slightly preferentially located in less disturbed regions with lower $G$ values. The optically redder sources are slightly 
less luminous in FIR than the bluer ones. All AGN ULIRGs have $G>0.55$. However these differences are not significant based
on the uncertainties of morphological parameter measurements, and the physical complexity of the merging systems.}
\label{other.relation}
\end{figure}

We test these hypotheses by plotting the $g-$band $Gini$ vs $M_{20}$ for the ULIRGs in Fig. \ref{other.relation}, 
with different symbol sizes representing their FIR luminosities and different symbol colors representing 
their $^{0.1}g-^{0.1}r$ colors.  There are no strong relations between the optical colors
and $G$ or $M_{20}$, except that optically redder sources are slightly more 
preferentially located in the non-merger regions with lower $G$ values.
This is further illustrated in the small panel in the same plot, where we plot the color distributions of sources 
with  $G\ge0.55$, the median $G$ of the whole sample, and of sources with $G<0.55$.  
The sources with lower $G$ values tend to distribute toward redder optical colors, although the reddest source 
has a higher-than-median $G$ value (0.62). All AGN ULIRGs have $G$ values greater than 0.5. 
There are no obvious correlations between the FIR luminosity and $G$ or $M_{20}$.

These results are broadly consistent with the scenario that the blue optical colors originate from more disturbed systems,
and these systems are not during the phase when the FIR emission peaks. However, as mentioned earlier, the 
uncertainties in measuring $G$ and $M_{20}$ could be substantial, and the measured morphologies are affected by orientation,
orbital parameters, viewing angle and dust content (Lotz et al. 2008b). We thus do not conclude any strong relations 
among the morphologies, the FIR luminosity, and the optical color of these system.

\section{SUMMARY}
\label{summary}

We present color-magnitude and morphological analysis of 54 ultraluminous infrared 
galaxies (ULIRGs) in the Sloan Digital Sky Survey (SDSS). The sample consists of all the ULIRGs from 
the IRAS 1Jy sample (Kim \& Sanders, 1998) that have been imaged in the Data Release 5 of SDSS, spanning a 
redshift range from 0.018 to 0.265 with a median redshift of 0.151.
The main results are:

$\bullet$ The ULIRGs are a very luminous group of galaxies in the optical, and the majority ($\sim93\%$) are more luminous than the 
median r-band absolute magnitude of the SDSS comparison galaxies in the same redshift range.

$\bullet$ The ULIRG sample forms a distinct group in the color-magnitude diagram. 24 out of the 52 unsaturated ULIRGs ($\sim46\%$)
lie outside the 90\% level contour of the SDSS galaxies in the same redshift range (NYU-VAGC, DR4; Blanton et al. 2005). 
The majority of the ULIRGs ($\sim87\%$)
have typical colors of the blue cloud, only $\sim6\%$ are located in the green valley, and $\sim7\%$ are located in the red sequence,
which implies that most of the ULIRGs are still undergoing copious star formation.

$\bullet$ There are 14 ULIRGs known to harbor an AGN (AGN ULIRGs). None of them are located in the green valley. 
We use two simple approaches to estimate the point source flux in these systems, and find that on average the central 
point source contributes less than one-third to the total luminosity in $r$ band. The host galaxies do not have a significantly 
different distribution in the color-magnitude diagram after the point sources removed.

$\bullet$ We perform $G$ and $M_{20}$ analysis on the ULIRGs and find that their distribution in the $G-M_{20}$ space is heterogeneous. 
Less than $\sim50\%$ (e.g. $\sim42\%$ in $g$ band) of the ULIRGs are located in the merger region found for local ULIRGs (LPM04). However the heterogeneous 
distribution is consistent with the morphology produced by numerical simulations (Lotz et al. 2008b), and we also discuss the uncertainty 
in the morphology measurements at the physical resolution of SDSS.


$\bullet$ We find that the ULIRGs have comparable optical luminosity as the SDSS QSOs within the same redshift
range but much redder. In comparison, the ULIRGs are much more luminous than the SDSS type 2 quasars within 
the same redshift range and much bluer. The distribution of the SDSS type 2 quasars peaks at the green valley. 
We perform two-dimensional K-S tests and the results show that the ULIRGs, the SDSS QSOs, and the SDSS type 2 
quasars are statistically different samples in the 
color-magnitude two-dimensional space.

$\bullet$ We study the relations among morphology, optical color, and far-infrared luminosity of the ULIRGs. These results 
are broadly consistent with the scenario that the blue optical colors originate from more disturbed systems, and these systems 
are not during the phase when the far-infrared emission peaks. However, the morphology parameters are
affected by the measurement uncertainties, as well as physical properties of the merger.  
We thus do not conclude any strong relations among morphology, far-infrared luminosity, and optical color of these system. 
 
Selected by their true infrared luminosity with no SED extrapolations, and with the uniformity of the SDSS optical photometry,
our ULIRG sample is a best low-redshift comparison sample to study the color-magnitude relation of high-redshift ULIRGs 
and sub-millimeter galaxies. Although the morphological study of our sample is limited by the physical resolution of SDSS, the 
size and structures of a galaxy might also be smaller at high redshift, and therefore our sample also provides a comparison
to study the morphology of high-redshift ULIRGs and sub-millimeter galaxies. We will discuss the color-magnitude relation
and morphology for a sample of the most luminous infrared galaxies at $z\sim1$ in a succeeding paper.

\acknowledgements

The authors would like to thank Dr. Jennifer Lotz for the use of the $G-M_{20}$ code. 
Yuxi Chen would like to thank Dr. Daniel McIntosh for inspiring comments and suggestions. We also appreciate
some useful discussions with Prof. David Sanders and Prof. Nicholas Scoville. 
This material is based upon work supported by the National Science Foundation under Grant No. NSF 0332504.

Funding for the creation and distribution of the SDSS Archive has been provided by the Alfred P. Sloan Foundation, 
the Participating Institutions, the National Aeronautics and Space Administration, the National Science Foundation, 
the U.S. Department of Energy, the Japanese Monbukagakusho, and the Max Planck Society. The SDSS Web site is http://www.sdss.org/.

The SDSS is managed by the Astrophysical Research Consortium (ARC) for the Participating Institutions. 
The Participating Institutions are The University of Chicago, Fermilab, the Institute for Advanced Study, the Japan Participation Group, 
The Johns Hopkins University, the Korean Scientist Group, Los Alamos National Laboratory, the Max-Planck-Institute for Astronomy (MPIA), 
the Max-Planck-Institute for Astrophysics (MPA), New Mexico State University, University of Pittsburgh, University of Portsmouth, 
Princeton University, the United States Naval Observatory, and the University of Washington.

\clearpage

\newpage
\begin{deluxetable}{lrrrrrrrrrrrrrccl}
\tablecaption{Sample Summary\label{tab:sample}}
\rotate
\tabletypesize{\scriptsize}
\tablecolumns{16}
\tablewidth{0pt}
\tablehead{
\colhead{Source}& \colhead{$z$} & \colhead{log$L_{FIR}$} & \colhead{$mag_u$} & \colhead{$u_{err}$}
& \colhead{$mag_g$} & \colhead{$g_{err}$} & \colhead{$mag_r$} & \colhead{$r_{err}$} & \colhead{$mag_i$} 
& \colhead{$i_{err}$} & \colhead{$mag_z$} & \colhead{$z_{err}$} & \colhead{$M_{^{0.1}r}$} & \colhead{$^{0.1}g-^{0.1}r$} & \colhead{2$r_{P,r}$}& \colhead{AGN} \\
\colhead{}&\colhead{}&\colhead{$\rm L_{\odot}$}&\colhead{(mag)}&\colhead{(mag)}&\colhead{(mag)}
& \colhead{(mag)}&\colhead{(mag)}&\colhead{(mag)}&\colhead{(mag)}&\colhead{(mag)}&\colhead{(mag)}&\colhead{(mag)}&\colhead{(mag)} 
&\colhead{(mag)}&\colhead{(\arcsec)}&\colhead{}\\
\colhead{(1)}&\colhead{(2)}&\colhead{(3)}&\colhead{(4)}&\colhead{(5)}&\colhead{(6)}&\colhead{(7)}&\colhead{(8)}
&\colhead{(9)}&\colhead{(10)}&\colhead{(11)}&\colhead{(12)}&\colhead{(13)}&\colhead{(14)}&\colhead{(15)}&\colhead{(16)}&\colhead{17}}
\startdata
FSC01166-0844&0.118&12.0&18.8&0.21&17.7&0.02&17.0&0.01&16.5&0.01&16.3&0.04&-20.9&0.66&14.9&
\\
FSC01572+0009&0.163&12.4&15.6&0.01&15.6&$<$0.01&15.3&$<$0.01&15.1&$<$0.01&15.0&0.01&-23.2&0.20&9.5&Sy1\tablenotemark{a,}\tablenotemark{b}
\\
FSC03209-0806&0.166&12.1&18.5&0.09&17.6&0.01&17.0&0.01&16.7&0.01&16.5&0.04&-21.6&0.50&12.3&
\\
FSCZ03521+0028&0.152&12.4&21.1&0.88&19.7&0.10&18.6&0.05&17.9&0.03&17.3&0.06&-19.9&1.00&14.9&
\\
FSC08201+2801&0.168&12.1&16.4&0.11&16.2&0.02&16.1&0.02&16.0&0.02&16.2&0.04&-22.5&0.04&14.6&
\\
FSC08572+3915&0.058&12.0&17.6&0.07&16.4&0.01&15.9&0.01&15.7&0.01&15.5&0.03&-20.3&0.52&20.0&
\\
FSC08591+5248&0.158&12.0&19.3&0.14&17.8&0.01&17.1&0.01&16.6&0.01&16.3&0.03&-21.5&0.66&8.8&
\\
FSC08474+1813&0.145&12.0&20.9&0.44&19.3&0.03&18.6&0.02&18.1&0.02&17.8&0.08&-19.8&0.70&6.7&
\\
FSC09039+0503&0.125&11.9&18.8&0.12&17.8&0.02&17.2&0.01&16.6&0.01&16.3&0.04&-20.8&0.60&14.6&
\\
FSC09116+0334&0.146&12.0&18.5&0.09&17.0&0.01&16.2&0.01&15.9&0.01&15.6&0.02&-22.1&0.68&16.2&
\\
FSC09539+0857&0.129&12.0&19.7&0.14&18.6&0.02&18.0&0.01&17.5&0.01&17.3&0.03&-20.1&0.59&6.6&
\\
FSC10035+2740&0.165&12.1&19.5&0.25&18.0&0.02&17.0&0.01&16.6&0.01&16.2&0.03&-21.7&0.84&13.3&
\\
FSC10091+4704&0.246&12.5&19.4&0.26&18.5&0.04&17.7&0.03&17.2&0.02&16.8&0.08&-22.1&0.66&29.2&
\\
FSC10190+1322&0.077&11.9&18.2&0.07&16.7&0.01&15.8&$<$0.01&15.3&$<$0.01&14.9&0.01&-21.0&0.86&18.2&
\\
FSC10378+1108&0.136&12.2&18.7&0.16&17.5&0.02&16.8&0.01&16.3&0.01&16.0&0.03&-21.4&0.63&18.2&
\\
FSC10494+4424&0.092&12.1&18.6&0.08&17.3&0.01&16.8&0.01&16.3&0.01&16.1&0.03&-20.4&0.48&15.2&
\\
FSC10594+3818&0.158&12.1&18.4&0.05&17.5&0.01&17.0&0.01&16.6&0.01&16.4&0.03&-21.6&0.49&7.0&
\\
FSC11028+3130&0.199&12.2&20.2&0.21&19.1&0.03&18.5&0.02&18.1&0.02&17.9&0.08&-20.6&0.45&6.6&
\\
FSC11119+3257&0.189&12.4&20.2&0.44&18.4&0.03&16.9&0.01&16.0&0.01&15.7&0.03&-22.3&1.36&11.7&Sy1\tablenotemark{d} 
\\
FSC11180+1623&0.166&12.1&19.4&0.20&18.4&0.03&17.6&0.08&17.2&0.02&16.9&0.02&-21.1&0.71&17.1&
\\
FSC11387+4116&0.149&12.0&19.9&0.19&18.3&0.02&17.4&0.01&16.9&0.01&16.6&0.03&-21.1&0.78&8.4&
\\
FSC11506+1331&0.127&12.2&18.8&0.09&17.8&0.02&17.3&0.01&16.8&0.01&16.7&0.03&-20.7&0.47&14.0&
\\
FSC11582+3020&0.223&12.4&20.1&0.29&19.0&0.04&18.1&0.02&17.7&0.02&17.8&0.11&-21.3&0.63&9.5&
\\
FSCZ11598-0112&0.151&12.3&17.1&0.03&16.8&0.01&16.4&0.01&16.0&0.01&15.9&0.02&-21.9&0.30&11.9&Sy1\tablenotemark{a}
\\
FSC12018+1941&0.168&12.3&19.4&0.16&18.1&0.02&17.5&0.01&17.1&0.01&17.0&0.04&-21.2&0.50&12.6&
\\
FSC12032+1707&0.217&12.4&19.1&0.17&18.2&0.02&17.6&0.02&17.0&0.01&17.1&0.06&-21.7&0.45&13.2&
\\
FSC12112+0305&0.073&12.2&17.2&0.06&16.2&0.01&15.6&0.01&15.2&0.01&15.0&0.03&-21.1&0.65&24.7&
\\
FSC12265+0219&0.159&12.3&12.5&$<$0.01&14.0&$<$0.01&13.8&$<$0.01&13.8&$<$0.01&12.7&$<$0.01&-24.7&0.27&7.9&Sy1\tablenotemark{b}
\\
FSC12447+3721&0.158&12.0&19.0&0.07&18.4&0.02&17.9&0.01&17.6&0.01&17.7&0.07&-20.6&0.35&7.5&
\\
FSC12540+5708&0.042&12.3&14.6&0.01&13.4&$<$0.01&13.3&$<$0.01&13.1&$<$0.01&12.4&$<$0.01&-22.1&0.15&20.2&Sy1\tablenotemark{b,c,d}
\\
FSC13218+0552&0.205&12.4&19.0&0.07&18.3&0.02&17.6&0.01&17.1&0.01&17.0&0.03&-21.6&0.56&7.6&Sy1\tablenotemark{d,e}
\\
FSC13342+3932&0.179&12.2&18.1&0.14&16.8&0.01&16.0&0.01&15.6&0.01&15.4&0.03&-22.9&0.61&29.3&Sy1\tablenotemark{d}
\\
FSC13428+5608&0.037&12.0&15.8&0.02&14.5&$<$0.01&14.0&$<$0.01&13.6&$<$0.01&13.4&0.01&-21.1&0.58&49.2&Sy2\tablenotemark{c,d}
\\
FSC13443+0802&0.135&12.0&17.9&0.08&16.7&0.01&16.0&0.01&15.5&0.01&15.3&0.02&-22.2&0.68&34.0&Sy2\tablenotemark{a}
\\
FSC13451+1232&0.122&12.0&18.0&0.07&16.4&0.01&15.4&$<$0.01&15.0&$<$0.01&14.8&0.01&-22.5&0.92&21.6&Sy2\tablenotemark{d}
\\
FSC13469+5833&0.158&12.1&18.9&0.14&17.8&0.02&17.1&0.01&16.7&0.01&16.5&0.05&-21.5&0.64&12.9&
\\
FSC13509+0442&0.136&12.1&22.2&1.09&18.2&0.02&17.5&0.01&17.1&0.01&16.8&0.06&-20.7&0.62&14.3&
\\
FSC13539+2920&0.108&11.9&18.6&0.08&17.5&0.01&16.9&0.01&16.3&0.01&16.2&0.03&-20.8&0.68&16.2&
\\
FSC14060+2919&0.117&11.9&17.6&0.04&16.7&0.01&16.3&0.01&15.9&$<$0.01&15.8&0.02&-21.5&0.40&15.5&
\\
FSC14070+0525&0.265&12.6&19.8&0.12&18.8&0.03&18.0&0.01&17.7&0.02&17.4&0.04&-21.8&0.46&7.3&Sy2\tablenotemark{d}
\\
FSC14121-0126&0.151&12.1&18.9&0.12&17.8&0.02&17.1&0.01&16.7&0.01&16.6&0.03&-21.4&0.65&19.9&
\\
FSC14197+0813&0.131&11.9&18.4&0.06&17.4&0.01&16.7&0.01&16.4&0.01&16.2&0.02&-21.3&0.58&7.4&
\\
FSC14202+2615&0.159&12.2&17.6&0.05&17.0&0.01&16.6&0.01&16.2&0.01&16.1&0.03&-21.9&0.33&15.1&
\\
FSC14394+5332&0.105&11.9&14.5&$<$0.01&16.5&0.01&15.9&0.01&15.4&$<$0.01&13.8&0.01&-21.7&0.58&30.4, 8.4\tablenotemark{\dag}&Sy2\tablenotemark{d}
\\
FSC15001+1433&0.162&12.3&18.1&0.14&17.2&0.02&16.7&0.01&16.3&0.01&16.1&0.04&-21.9&0.41&13.03, 8.59\tablenotemark{\dag}&Sy2\tablenotemark{d}
\\
FSC15043+5754&0.151&12.0&19.0&0.07&18.2&0.01&17.6&0.01&17.2&0.01&17.0&0.04&-20.8&0.47&7.3&
\\
FSC15206+3342&0.125&12.0&17.3&0.01&16.8&$<$0.01&16.4&$<$0.01&15.9&$<$0.01&16.1&0.01&-21.5&0.31&6.7&
\\
FSC15225+2350&0.139&12.0&19.0&0.12&18.2&0.02&17.4&0.01&16.9&0.01&16.8&0.05&-20.9&0.72&13.0&
\\
FSC15327+2340&0.018&12.1&15.2&0.02&13.6&$<$0.01&12.9&$<$0.01&12.5&$<$0.01&12.3&$<$0.01&-20.5&0.76&66.6&
\\
FSC16300+1558&0.242&12.6&18.9&0.13&18.3&0.03&17.4&0.01&17.1&0.01&16.9&0.06&-22.2&0.57&10.8&
\\
FSC16333+4630&0.191&12.3&18.9&0.09&18.0&0.02&17.3&0.01&17.0&0.01&16.9&0.05&-21.6&0.48&13.9&
\\
FSC16468+5200&0.150&11.9&19.0&0.12&18.1&0.02&17.5&0.01&17.2&0.01&17.2&0.07&-20.9&0.52&12.4&
\\
FSC16474+3430&0.111&12.0&17.5&0.04&16.7&0.01&16.2&$<$0.01&15.8&$<$0.01&15.6&0.02&-21.4&0.40&11.5&
\\
FSC17179+5444&0.147&12.1&19.1&0.12&17.6&0.01&16.9&0.01&16.3&0.01&16.0&0.02&-21.5&0.68&13.5&Sy2\tablenotemark{a}
\enddata
\tablecomments{(1): $IRAS$ Faint Source Catalog name (Moshir et al. 1992). (2): Optical redshifts (Kim \& Sanders 1998). 
(3): FIR luminosity (from 60$\mu$m and 100$\mu$m IRAS fluxes; Kim \& Sanders 1998). 
(4)-(13): Photometry results, this work (see text). (14): Absolute magnitude in $^{0.1}r$ (see text). 
(15): $^{0.1}g-^{0.1}r$ color (see text). (16): Two Petrosian radii in $r-$band adopted as the photometry aperture (see text). 
\dag: For the sources with multiple parts well separated, the apertures for all parts are shown.
(17): AGN types by previous optical and near-infrared spectroscopic studies.
Sy1 and Sy2 stand for Seyfert 1 and Seyfert 2 types, respectively. 
References are: [a] VKS99; [b] V\'{e}ron-Cetty \& V\'{e}ron 1989; 
[c] Veilleux et al. 1995; [d] Kim, Veilleux \& Sanders 1998; [e] Low et al. 1998.}
\end{deluxetable}
\clearpage

\newpage
\begin{deluxetable}{cccccc}
\tabletypesize{\scriptsize}
\tablewidth{0pt}
\tablecaption{Photometry Statistics of the ULIRGs and SDSS Comparison Sample \label{statistics}}
\tablehead{&\multicolumn{2}{c}{ULIRGs} &\multicolumn{3}{c}{SDSS Comparison Sample}}
\startdata
  Number of Sources & \multicolumn{2}{c}{52\tablenotemark{[1]}}&\multicolumn{3}{c}{436762} \\
 $\langle M_{^{0.1}r}\rangle \pm\sigma_{M_{^{0.1}r}}$ &\multicolumn{2}{c}{$-21.4\pm0.72$} &\multicolumn{3}{c}{$-20.2\pm1.17$}  \\
 $\langle ^{0.1}g-^{0.1}r\rangle \pm\sigma_{^{0.1}g-^{0.1}r}$ & \multicolumn{2}{c}{$0.57\pm0.22$}&\multicolumn{3}{c}{$0.77\pm0.25$} \\
 \\
 &   AGN ULIRGs&non-AGN ULIRGs & Red Sequence & Green Valley & Blue Cloud  \\
  \cline{1-6}
  Number of Sources & 12\tablenotemark{[1]}&40 & 216648 & 51904 & 168119 \\
  $\langle M_{^{0.1}r}\rangle \pm\sigma_{M_{^{0.1}r}}$ &$-22.1\pm0.60$ & $-21.1\pm0.62$ & $-20.6\pm1.01$ & $-20.4\pm1.01$ & $-19.8\pm1.26$  \\
   $\langle ^{0.1}g-^{0.1}r\rangle \pm\sigma_{^{0.1}g-^{0.1}r}$ & $0.61\pm0.30$ & $0.58\pm0.17$ & $0.95\pm0.15$ & $0.76\pm0.04$ & $0.52\pm0.18$
\enddata
\tablenotetext{[1]}{ The ULIRGs sample excludes the two saturated sources FSC12540+5708 and FSC12265+0219, and these two sources are also AGN ULIRGs.}
\end{deluxetable}
\clearpage

\newpage
\begin{deluxetable}{lccccccccccc}
\tabletypesize{\scriptsize}
\rotate
\tablewidth{0pt}
\tablecaption{Two Dimensional Kolmogorov-Smirnov Test for $^{0.1}g-^{0.1}r$ and $M_{^{0.1}r}$ \label{kstest}}
\tablehead{ & Number & \multicolumn{2}{c}{SDSS QSOs} & \multicolumn{2}{c}{Type 2 quasars}& \multicolumn{2}{c}{ULIRGs}
 & \multicolumn{2}{c}{AGN ULIRGs} & \multicolumn{2}{c}{non-AGN ULIRGs}  \\
 \cline{3-12}
 & & $D$\tablenotemark{[1]} & $P(>Z)$\tablenotemark{[2]} &$D$ &$P(>Z)$ & $D$ & $P(>Z)$ & $D$ & $P(>Z)$ & $D$ & $P(>Z)$ }
\startdata 
 SDSS QSOs & 1170 & ...&... & 0.86& 0.00& 0.59 & $1.52\times10^{-12}$ & 0.79 & $2.70\times10^{-7}$ & 0.62 & $8.31\times10^{-11}$ \\
 Type 2 quasars & 402 & 0.86 & 0.00&...&...&0.69&$1.26\times10^{-15}$& 0.87 & $1.87\times10^{-8}$ &0.66 & $2.11\times10^{-11}$\\
 ULIRGs & 54 & 0.59 & $1.52\times10^{-12}$ & 0.69 & $1.26\times10^{-15}$ &...&...& 0.49 & $1.17\times10^{-2}$ & 0.17 & 6.13$\times10^{-1}$\\
 AGN ULIRGs &14&0.79&$2.70\times10^{-7}$ & 0.87 & $1.87\times10^{-8}$ & 0.49 & $1.17\times10^{-2}$ &... &...& 0.66 & $2.82\times10^{-4}$  \\
  non-AGN ULIRGs &40&0.62&$8.31\times10^{-11}$ & 0.66& $2.11\times10^{-11}$&0.17 & 6.13$\times10^{-1}$& 0.66 & $2.82\times10^{-4}$&...&...  \\
   \\
  Red Sequence&216648&0.98&0.00& 0.60 & 0.0 & 0.94&$6.05\times10^{-32}$&0.91&$1.18\times10^{-9}$&0.94&$2.24\times10^{-24}$\\
  Blue Cloud&220023&0.87&0.00 & 0.43 & $9.81\times10^{-45}$& 0.64&$1.47\times10^{-15}$&0.80&$1.57\times10^{-7}$ &0.62&$5.62\times10^{-11}$
\enddata
\tablenotetext{[1]}{ $D$ is the maximum absolute difference between the two samples' cumulative distribution functions.
In two dimensional KS test we considered all four possible ranking combinations. For details see Peacock (1983).}
\tablenotetext{[2]}{ $Z=\sqrt{\frac{N_1N_2}{N_1+N_2}}D$ is the test statistics,
where $N_1$ and $N_2$ are the sample sizes. $P(>Z)$ gives the probability of
{\it not rejecting} the null hypothesis $H_0$: the two samples are drawn from the same population. The calculation of $P(>Z)$ utilize
the function {\sffamily probks} in the {\it Numerical Recipes} (Press et al. 1992).}
\label{KStable}
\end{deluxetable}
\clearpage

\end{document}